\documentclass{emulateapj}
\usepackage{natbib}
\usepackage{amsmath}

\newcommand{\lsim}{\mbox{$_<\atop^{\sim}$}}

\newcommand{\ha}{H$\alpha$}
\newcommand{\hb}{H$\beta$}
\newcommand{\oiii}{[O{\small III}]}
\newcommand{\oii}{[O{\small II}]}
\newcommand{\nii}{[N{\small II]}}

\begin{document}

\title{Predicting the redshift 2 \ha\ luminosity function using \oiii\ emission line galaxies}
\author{Vihang Mehta\altaffilmark{1}, 
Claudia Scarlata\altaffilmark{1}, 
James W. Colbert\altaffilmark{2}, 
Y. S. Dai\altaffilmark{3}, 
Alan Dressler\altaffilmark{4},
Alaina Henry\altaffilmark{5}, 
Matt Malkan\altaffilmark{6}, 
Marc Rafelski\altaffilmark{5},
Brian Siana\altaffilmark{7}, 
Harry I. Teplitz\altaffilmark{3}, 
Micaela Bagley\altaffilmark{1},
Melanie Beck\altaffilmark{1}, 
Nathaniel R. Ross\altaffilmark{6}, 
Michael Rutkowski\altaffilmark{1}, 
Yun Wang\altaffilmark{3,8}}

\altaffiltext{1}{Minnesota Institute for Astrophysics, University of Minnesota, Minneapolis, MN 55455, USA} 
\altaffiltext{2}{Spitzer Science Center, California Institute of Technology, Pasadena, CA 91125, USA}
\altaffiltext{3}{Infrared Processing and Analysis Center, California Institute of Technology, Pasadena, CA 91125, USA}
\altaffiltext{4}{Observatories of the Carnegie Institution for Science, Pasadena, CA 91101, USA}
\altaffiltext{5}{Astrophysics Science Division, Goddard Space Flight Center, Code 665, Greenbelt, MD 20771, USA}
\altaffiltext{6}{Department of Physics \& Astronomy, University of California, Los Angeles, CA 90095, USA}
\altaffiltext{7}{Department of Physics and Astronomy, University of California Riverside, Riverside, CA 92521, USA}
\altaffiltext{8}{Homer L. Dodge Department of Physics \& Astronomy, University of Oklahoma, Norman, OK 73019, USA}

\begin{abstract}

Upcoming space-based surveys such as \textit{Euclid} and \textit{WFIRST-AFTA} plan to measure Baryonic Acoustic Oscillations (BAOs) in order to study dark energy.  These surveys will use IR slitless grism spectroscopy to measure redshifts of a large number of galaxies over a significant redshift range. In this paper, we use the WFC3 Infrared Spectroscopic Parallel Survey (WISP) to estimate the expected number of \ha\ emitters observable by these future surveys. WISP is an ongoing \textit{Hubble Space Telescope} slitless spectroscopic survey, covering the 0.8 -- 1.65$\mu$m wavelength range and allowing the detection of \ha\ emitters up to $z\sim1.5$ and \oiii\ emitters to $z\sim2.3$. We derive the \ha --\oiii\ bivariate line luminosity function for WISP galaxies at $z\sim1$ using a maximum likelihood estimator that properly accounts for uncertainties in line luminosity measurement, and demonstrate how it can be used to derive the \ha\ luminosity function from exclusively fitting \oiii\ data. Using the $z\sim2$ \oiii\ line luminosity function, and assuming that the relation between \ha\ and \oiii\ luminosity does not change significantly over the redshift range, we predict the \ha\ number counts at $z\sim2$ -- the upper end of the redshift range of interest for the future surveys. For the redshift range $0.7<z<2$, we expect $\sim$3000 galaxies/deg$^2$ for a flux limit of $3 \times 10^{-16}$ ergs s$^{-1}$ cm$^{-2}$ (the proposed depth of \textit{Euclid} galaxy redshift survey) and $\sim$20,000 galaxies/deg$^2$ for a flux limit of $\sim$10$^{-16}$ ergs s$^{-1}$ cm$^{-2}$ (the baseline depth of \textit{WFIRST} galaxy redshift survey).

\end{abstract}

\section{Introduction}
\label{sec:intro}

The origin of dark energy, responsible for the accelerated expansion of the universe first observed by \citet{riess98} and \citet{perlmutter99}, is one of the most important unsolved problems in cosmology today, and significant effort is being devoted to constrain its properties. Dark energy affects both the expansion history of the Universe as well as the growth of structures. Both effects can be observationally constrained through large galaxy redshift surveys, which enable the measurement of Baryon Acoustic Oscillations (BAOs; thus, constraining the cosmic expansion history) and large scale redshift-space distortions (thus, constraining the growth history of the large scale structure). The combination of these two measurements allows the differentiation between an unknown energy component and modification of general relativity as the cause of observed cosmic acceleration \citep{guzzo08, wang08}. 

Upcoming space-based missions, ESA's \textit{Euclid} \citep{laureijs12} and NASA's \textit{WFIRST-AFTA} \citep{green12,dressler12,spergel15}, will perform complementary galaxy redshift surveys to map the large-scale structure and its evolution over a cosmic time covering the last 10 billion years. Both \textit{Euclid} and \textit{WFIRST-AFTA} will use the \ha\ $\lambda$6563 line and \oiii\ $\lambda \lambda$4959+5007 doublet to select emission line galaxies as tracers of the large scale structure at 0.7\lsim $z$\lsim 2 (in \ha ) and 2\lsim $z$\lsim 2.7 (in \oiii ). The performance of the planned missions can be quantified by a figure-of-merit, which describes their ability to measure the present value and time evolution of the dark energy equation of state. The figure-of-merit for a dark energy survey depends on the number density of tracer galaxies available at each redshift. It is therefore critical to have a reliable and sufficiently precise knowledge of the expected number of \ha\ and \oiii\ galaxies in the survey volumes.

\textit{Euclid} and \textit{WFIRST-AFTA} will both perform IR slitless spectroscopy of emission line galaxies, in a way similar (scaled by many orders of magnitude in area) to the IR spectroscopic surveys that are being conducted with the  Wide Field Camera 3 (WFC3) on-board the \textit{Hubble Space Telescope} (HST). The WFC3 Infrared Spectroscopic Parallel Survey \citep[WISP;][]{atek10} is an on-going pure-parallel near-infrared grism spectroscopic survey using the WFC3 camera. While covering a substantially smaller area, WISP is very similar in many respects to the planned dark--energy surveys, and thus, can be used to test number count predictions, redshift measurement accuracy, target selection function, as well as completeness for the planned surveys. Towards this goal, \cite{colbert13} predicted the number counts of \ha\ -- emitting galaxies in the redshift range $0.3<z<1.5$. In this paper we extend the work by \cite{colbert13} and estimate \ha\ number counts out to $z\sim2$.

Ground-based wide-field narrow-band surveys like HiZELS \citep{geach10,sobral09,sobral12,sobral13} and NEWFIRM \ha\ Survey \citep{ly11} have been able to measure the \ha\ luminosity function in the redshift range of interest ($0.7 \lsim z \lsim 2$). However, while having the advantage of high sensitivity to emission lines and covering significant areas in the sky, these surveys can only map very narrow redshift ranges. Volume densities of galaxies can thus be strongly affected by the presence of large scale structures in the field.   Moreover, samples selected with narrow band surveys, without an extensive spectroscopic followup, can suffer from contamination by emission lines at different redshifts \cite[e.g., \oiii\ and \oii;][]{martin08,henry12}. Finally, even narrow-band surveys with multiple filters tuned to identify multiple emission lines at the same redshifts still rely on continuum detections and miss the lowest mass galaxies, to which WISP is very sensitive.

WISP's grism coverage includes \ha\ for the redshift range $0.3<z<1.5$ and \oiii\ for $0.7<z<2.3$. Since \ha\ is not directly covered by WISP at $z>1.5$, we cannot measure the \ha\ luminosity function up to z$\sim$2 explicitly. However, we can use the \oiii\ coverage to estimate the \ha\ luminosity function and number counts. In order to do this, we compute the bivariate \ha --\oiii\ line luminosity function in the redshift where both lines are visible, and use the resulting fit at higher redshifts, where only \oiii-emitters are observable in the WISP data. We also introduce a modified Maximum Likelihood Estimator to obtain the best-fit model parameters, which accounts for measurement uncertainties in the line luminosity (which can substantially affect the shape of the bright end of the luminosity function).

The paper is organized as follows: in Section~\ref{sec:data}, we summarize the new WISP data that we use in this work; Section~\ref{sec:ha_oiii_trend} discusses the observed \ha --\oiii\ relation; and Section~\ref{sec:blf_def} describes the parametrization for the bivariate luminosity function. In Section~\ref{sec:mle}, we discuss the Maximum Likelihood Estimator modified to account for uncertainties in the line flux, and the fitting procedure, which we use to derive the \ha --\oiii\ bivariate luminosity function at redshift $z\sim1$ and the result is discussed in Section~\ref{sec:blf}. Further, in Section~\ref{sec:z1}, we demonstrate the ability to recover the \ha\ luminosity function as well as number counts from fitting only the \oiii\ data at redshift $z\sim1$. Lastly, we fit the \oiii\ luminosity function and use it to derive the \ha\ luminosity function and number counts at redshift $z\sim2$ in Section~\ref{sec:z2}, along with the final number count estimates for the upcoming dark-energy surveys.

Throughout this paper, we assume standard cosmology with $\Omega_m=0.3$, $\Omega_\lambda = 0.7$ and $H_0=70$ km s$^{-1}$ Mpc$^{-1}$.

\section{Data}
\label{sec:data}

The WFC3 Infrared Spectroscopic Parallel Survey (WISP) is discussed in full detail in \cite{atek10}. Briefly, WISP consists of \textit{HST} WFC3 pure--parallel IR slitless spectroscopic observations and imaging of hundreds of uncorrelated high-latitude fields. The spectroscopy is performed using the  G102 ($0.8-1.15$ $\mu$m, R $\sim$ 210) and G141 ($1.15-1.65$ $\mu$m, R $\sim$ 130) grisms, while the associated near-IR imaging is obtained with the F110W and F160W, filters. For this paper, we use data from 52 separate fields for which both G102 and G141 grism spectroscopy are available, covering a total of 182 arcmin$^2$. These fields include 23 new WISP fields in addition to the ones used by \cite{colbert13}. All data are processed with a combination of the WFC3 pipeline CALWF3 and custom scripts, to account for the lack of dithering of the pure parallel data \citep[see][]{atek10}. The siltless extraction package aXe 2.0 \citep{kummel09} is used to perform the spectral extraction. We perform a blind search for emission lines in all fields (both grisms) down to a typical $5\sigma$ line flux-limit of $(3-5) \times 10^{-17}$ erg s$^{-1}$ cm$^{-2}$ as explained in Ross et al. (2015, in prep). In order to remove the high contamination rate from false and/or spurious sources due to the parallel, slitless nature of the WISP survey, every candidate emission line undergoes independent visual inspection by two team members. This process of visually confirming the emission lines is described in further detail in \cite{colbert13}.

In this work, we are interested in the \ha\ $\lambda$6563 line and \oiii\ $\lambda \lambda$4959+5007 doublet, which are covered by WISP survey over the redshift ranges: $0.3<z<0.7$ (only \ha), $0.7<z<1.5$ (both \ha +\oiii ), $1.5<z<2.3$ (only \oiii ). We exclude all sources with any ambiguity in their redshift determination among multiple reviewers. Specifically, we only retain sources with a quality flag $<16$, which implies consensus among the independent reviewers (Ross et al. 2015, in prep). Since \ha\ line and \nii\  $\lambda \lambda$6548+6584 doublet is not resolved in the WISP grisms, we apply a correction factor of 0.71 to the \ha\ luminosities to account for \nii\ contamination, similar to \cite{colbert13}. Although \cite{villar08} and \cite{cowie11} report decreasing \nii /\ha\ ratio with increasing \ha\ equivalent width, this ratio is nearly constant up to \ha\ EW $\sim$200\AA, above which the correlation steepens. The fraction of galaxies in our sample for which we may overestimate the \nii\ contribution due to the assumed constant correction is only 10\%.

In what follows, we use the completeness analysis from \cite{colbert13}, who performed extensive simulations to quantify the survey incompleteness as a function of line signal--to--noise ratio, equivalent width (EW), as well as galaxy size. The completeness simulations followed the full line extraction process after adding artificial sources to the real data, spanning a range in redshifts, radii, brightnesses, equivalent widths (EWs), as well as using different empirical spectral templates from the Kinney-Calzetti Altas. These simulations show that slitless spectroscopic surveys display some level of incompleteness even at large EWs and line fluxes, primarily because of spectral overlap and line mis-identification.

\section{\ha --\oiii\ Trend}
\label{sec:ha_oiii_trend}

\begin{figure}[!t]
\includegraphics[width=0.46\textwidth]{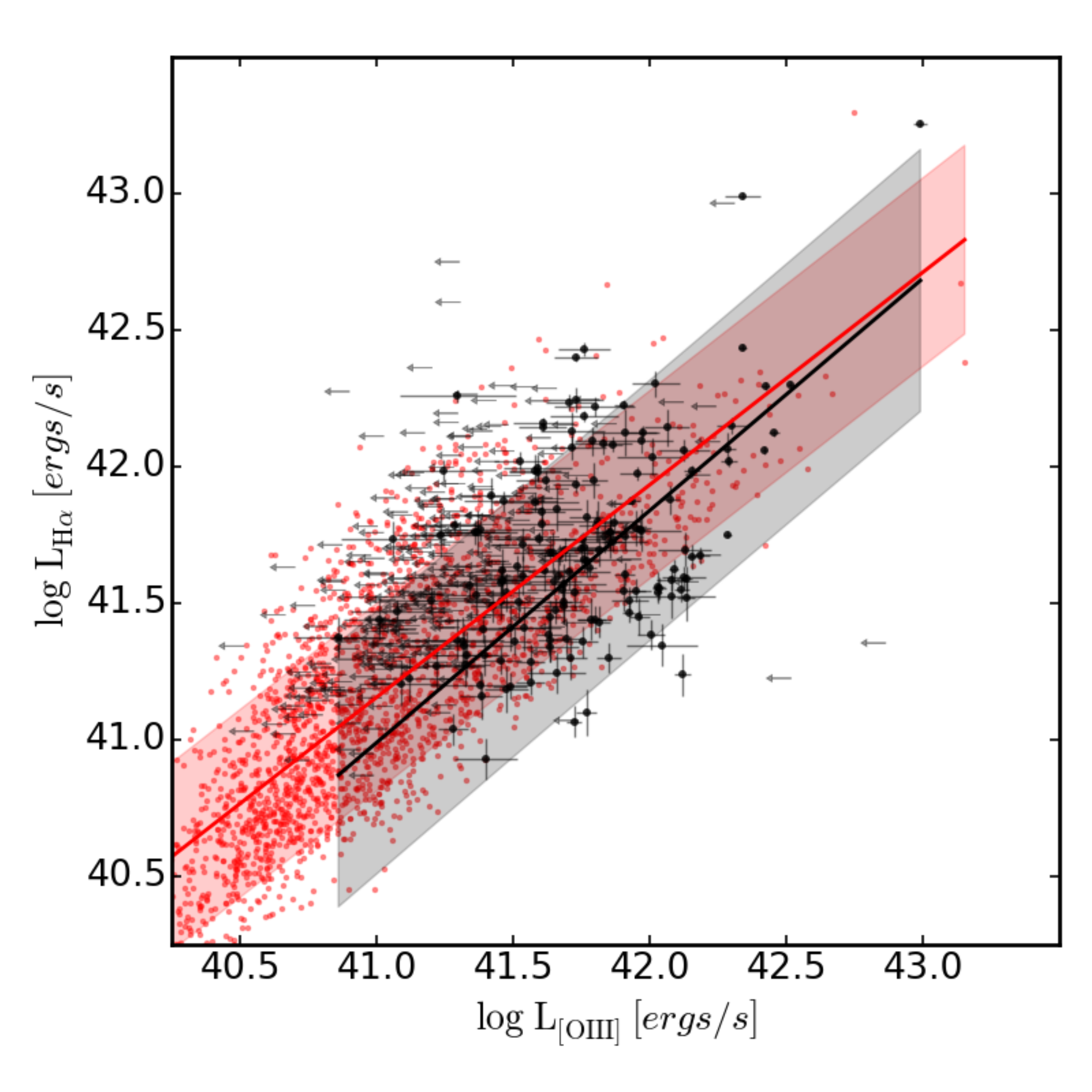}
\caption{A broad correlation is observed between \ha\ and \oiii\ luminosity in the SDSS star-forming galaxies in $0.2<z<0.3$ (\textit{red points}) as well as our WISP sample in redshift range $0.8<z<1.2$ (\textit{black points}). The \textit{black} and \textit{red} lines are linear fits to the WISP and SDSS data, respectively and the shaded regions show the $1\sigma$ deviations from the linear fit ($\sim0.5$ dex).}
\label{fig:sample_blf}
\end{figure}

Figure~\ref{fig:sample_blf} shows the observed \ha\ against \oiii\ line luminosity, for the WISP galaxies in the redshift range $0.8<z<1.2$. The \textit{red} points in Figure~\ref{fig:sample_blf} show line luminosities for a sample of 2141 star-forming galaxies in the Sloan Digital Sky Survey \citep{thomas13}, limited to the redshift range $0.2<z<0.3$, in order to ensure that $3''$ spectroscopic aperture contains most of the galaxy flux (rather than just the central nucleus), and to avoid evolutionary effects.

Despite the large scatter (on the order of 0.5 dex), the two line luminosities are broadly correlated, both in the SDSS as well as WISP samples. This is not surprising: both \ha\ and \oiii\ are  observed in the ionized gas in star-forming galaxies, although, while the \ha\ luminosity scales directly with the ionizing fluxes of embedded young, hot stars, the \oiii\ luminosity is more strongly dependent on variations in the oxygen excitation state, overall gas oxygen abundance, gas density, as well as dust reddening \citep[e.g.,][]{kennicutt92,moustakas06}. Figure~\ref{fig:sample_blf} also shows that the \ha --\oiii\ relation does not evolve significantly in the $\sim$4.5 billion years elapsed between $z\sim0.25$ (SDSS galaxies) and $z\sim1$ (WISP galaxies). Although the gas oxygen abundance is observed to evolve with cosmic time, the similarity between the observed trends suggests that any evolution is masked by the large scatter introduced by the range of physical conditions present in galaxies.

As we will show in Section~\ref{sec:z1}, the observed broad correlation between \ha\ and \oiii\ luminosity is sufficient for the goal of estimating the number of \ha\ emitters from the numbers of \oiii\ emitters, as long as the scatter is appropriately taken into account.  In computing the number of \ha\ emitters from \oiii\ emitters at higher redshifts, we will assume that the trend between \ha\ and \oiii\ does not change in the $\sim$2.2 billion years between $z\sim 1$ and $z\sim 1.8$.

\section{Parametrization of the Bivariate Line luminosity function}
\label{sec:blf_def}

The goal of this work is to predict the \ha\ number counts at redshift $z\sim2$ using the WISP dataset. At this redshift, the WISP survey does not cover \ha\ directly, but it does cover the \oiii\ $\lambda \lambda$4959+5007 doublet. Hence, we estimate the \ha\ number counts from the available \oiii\ number counts. In order to do this, we start by computing the \ha --\oiii\ bivariate line luminosity function (LLF), which describes the volume density of sources as a function of both the \ha\ and \oiii\ luminosities. 

The most widely used parametric form for galaxy luminosity functions is the Schechter function \citep{schechter76}, which is fully described by the parameters $L^\star$ (characteristic luminosity), $\phi^\star$ (number of galaxies per unit volume at $L^\star$), and $\alpha$ (faint end slope). This function is found to reproduce the LLF of both \oiii\ and \ha-selected WISP galaxies \citep{colbert13}.

We define the bivariate LLF by combining a Schechter form (to describe the \oiii\ LLF) with the conditional probability for finding an \ha\ source given an \oiii\ luminosity\footnote{In this definition, the marginalized function over  all \oiii\ luminosities is not an exact Schechter form -- but very close to it.}. Thus, we parametrize the \oiii\ LLF as:

\begin{equation} \label{eqn:LLF}
\psi( \mathrm{L_{OIII}}) \ d\mathrm{L_{OIII}} =\phi^\star
\left( \frac{\mathrm{L_{OIII}}}{L^\star} \right) ^{\alpha} \mathrm{exp}\left[- \frac{\mathrm{L_{OIII}}}{L^\star} \right] \frac{d\mathrm{L_{OIII}}}{L^\star}
\end{equation}

\noindent where $\mathrm{L_{OIII}}$ is the \oiii\ line luminosity.  We adopt a log-normal distribution to describe the conditional probability that a galaxy with \oiii\ luminosity in the range ($\mathrm{L_{OIII}}$, $\mathrm{L_{OIII}}$+$d\mathrm{L_{OIII}}$) has \ha\ luminosity in the range ($\mathrm{L_{H\alpha}}$, $\mathrm{L_{H\alpha}}$+$d\mathrm{L_{H\alpha}}$):

\begin{equation}\label{eqn:CProb}
\begin{split}
p(\mathrm{ L_{H\alpha} | L_{[OIII]}}) \ d\mathrm{L_{H\alpha}} = & \\
\frac{1}{\sigma_\mathrm{lnL_{H\alpha}} \sqrt{2\pi}} & \cdot \mathrm{exp} \left[ - \frac{ \mathrm{ln}^2 (\mathrm{L_{H\alpha}} / \mathrm{ \langle L_{H\alpha} \rangle} ) } {2\sigma_\mathrm{lnL_{H\alpha}}^2} \right] \ \frac{d\mathrm{L_{H\alpha}}}{\mathrm{L_{H\alpha}}}
\end{split}
\end{equation}

\noindent where $\mathrm{\langle L_{H\alpha} \rangle}$ defines the mean expected \ha\ luminosity for a given \oiii, and $\sigma_\mathrm{lnL_{H\alpha}}$ is the scatter around the mean relation. $\mathrm{\langle L_{H\alpha} \rangle}$ and $\mathrm{L_{OIII}}$ are related through the ratio $r$, such that:

\begin{equation} \label{eqn:rel}
\frac{\mathrm{\langle L_{H\alpha} \rangle}}{\mathrm{L_0}} = r \cdot \left( \frac{\mathrm{L_{OIII}}} {\mathrm{L_0}} \right) ^{\beta}
\end{equation}

The ratio $r$ is defined as the expected $\mathrm{L_{H\alpha}} / \mathrm{L_0}$ at a nominal luminosity $\mathrm{L_0}$, where we arbitrarily choose $\mathrm{L_0} = 10^{40}$ ergs s$^{-1}$. The LLF and conditional probability equations (Equations \ref{eqn:LLF} and \ref{eqn:CProb}, respectively) can now be combined into the bivariate luminosity function, expressed in terms of  the \ha\ and \oiii\ $\mathrm{log_{10}}$ luminosities, ($x$ and $y$ respectively) as:

\begin{equation}\label{eqn:BLF}
\begin{split}
\Psi(x,y; \vec{\textbf{P}}) & \ \mathrm{d}x \ \mathrm{d}y= 
\ \mathrm{ln} 10 \cdot \left(\displaystyle \frac{10^y}{L^{\star}} \right)^{\alpha+1} \mathrm{exp}\left[-\frac{10^y}{L^{\star}} \right] \cdot \\
&\frac{\mathrm{ln}10}{\sigma_\mathrm{lnL_{H\alpha}}\sqrt{2\pi}}  \cdot \mathrm{exp}\left[{\frac{
{-[ x - \langle x \rangle]}^{2}}{2 (\sigma_\mathrm{lnL_{H\alpha}} / \mathrm{ln} 10)^2}}\right] \mathrm{d} x \ \mathrm{d} y
\end{split}
\end{equation}

\noindent where $\langle x \rangle$ is defined by Equation~\ref{eqn:rel} as,

\begin{equation}
\langle x \rangle - \mathrm{log \ L_0} = \mathrm{log} \ r + \beta (y - \mathrm{log \ L_0})
\end{equation}

The bivariate LLF, $\Psi(x,y; \vec{\textbf{P}})$, in Equation~\ref{eqn:BLF} can now be fully described by the set of parameters $\vec{\textbf{P}} = [\alpha,L^{\star},\beta,r,\sigma_\mathrm{lnL_{H\alpha}}]$. The formulation for the bivariate luminosity function described here, has been partly inspired from the size-luminosity bivariate distribution from \cite{huang13}.

\section{Fitting Procedure}
\label{sec:mle}

There are various parametric as well as non-parametric techniques used to derive the best fit parameters of luminosity functions (LF); to name a few: the $V_{max}$ estimator by \citet{trumpler53}, the $C^-$ method by \citet{lynden71}, the maximum likelihood estimator by \citet[][hereafter STY]{sty79}, and the stepwise maximum likelihood estimator by \citet{efstathiou88}.  In this paper, we use the STY parametric maximum likelihood estimator (MLE), modified to account for uncertainties in the measurements of the line luminosity, as explained in Section~\ref{subsec:modified_mle}. One of the major advantages of the MLE is that it allows us to fit the data without binning. Particularly for small samples, this technique reduces the biases introduced by the choice of bin-size or bin-center as well as any effects due to changing completeness and effective volume within the bin \citep[e.g.,][]{maiz05}. The modification of the method we introduce in Section~\ref{subsec:modified_mle} allows us to account for significant measurement uncertainties on the data. Large photometric uncertainties can impact the determination of the best fit parameters of the LLF, particularly at the bright end, where the number density of galaxies is a steep function of galaxy's luminosity \citep{henry12}. In slitless spectroscopy, photometric uncertainties can be large even for bright galaxies (i.e., the noise is not only due to the sky background but also to the possible contamination of the line flux due to the continuum of overlapping spectra), so it is crucial to account for line luminosity uncertainty in the fitting process.

\subsection{Original MLE}
\label{subsec:original_mle}

The original MLE is a parametric estimator, where the best fit parameters are obtained by maximizing the likelihood function ($\mathcal{L}$) of observing the galaxy sample with respect to the parameters of the model. For a given LF parametric description $\Psi(L)$, the probability for detecting a given galaxy with log luminosity $L$ is given by:

\begin{equation}
P(L_{i}) = \frac{\Psi(L_{i}) \cdot V_{\mathrm{eff}}(L_{i})}{ \displaystyle \int \limits _{L_{lim}} ^\infty \Psi(L) \cdot V_{\mathrm{eff}}(L) \cdot \mathrm{d}L}
\end{equation}
\noindent where $V_{\mathrm{eff}}(L)$ is the effective volume of the survey. The effective volume varies with the galaxy's line luminosity and redshift, and can be written as:

\begin{equation} \label{eqn:veff}
V_{\mathrm{eff}}(L_{i}) = \displaystyle \int \limits _{z _{min}}^{z _{max}} \frac{dV_{comov}}{dz \cdot d\Omega}(z) \cdot C\left( L_{i},z \right) \Omega (z) \cdot dz
\end{equation}
\noindent where $[z_{min}, z_{max}]$ are the redshift range of the survey, $dV_{comov}/{dz \cdot d\Omega}$ is the differential co-moving volume at redshift $z$, $C(L,z)$ is the completeness function, and $\Omega(z)$ is the solid angle covered by the survey. The likelihood function for the full sample can then be computed as the product of the individual probabilities for all galaxies in the sample:

\begin{equation}
\mathcal{L} = \prod \limits_{i=1}^N P(L_i)
\end{equation}

The best fit parameters of the LF can be found by maximizing the likelihood function with respect to the model parameters. It is mathematically and computationally simpler to maximize the log-likelihood function:

\begin{equation}\label{eqn:loglike}
\mathrm{ln}\mathcal{L} = \sum \limits _{i=1}^{N} \mathrm{ln} P(L_i)
\end{equation}

Because this method involves ratios between the differential and integrated luminosity functions, the normalization ($\phi^\star$) cancels out and, hence, it cannot be determined by this likelihood maximization procedure. $\phi^\star$ can be computed following \citep[e.g.,][]{alavi14}:

\begin{equation}
\phi^{\star} = \frac{N}{ \displaystyle \int \limits _{L_{lim}}^{\infty} \Psi(L) \cdot V_{\mathrm{eff}}(L) \cdot \mathrm{d}L}
\end{equation}
\noindent where $N$ is the total number of sources in the sample and the survey incompleteness is accounted for by the effective volume.

\subsection{Modified MLE}
\label{subsec:modified_mle}

All astronomical observations have an associated measurement uncertainty. It is crucial to account for these uncertainties, particularly when fitting models that vary steeply as a function of the independent variable (e.g., at the bright end of the Schechter function). In such cases, the best fit-parameters can change significantly, if even a few sources are scattered toward or away from the bright end due to photometric uncertainties. This problem is particularly important for slitless spectroscopic data, where line flux uncertainties can be substantial even for bright line fluxes, due to the common overlapping of spectral traces.  We modify the original prescription to account for observational uncertainties as follows. 

Instead of calculating the probability that a galaxy is exactly at a given luminosity, we marginalize the Schechter function over the luminosity error probability distribution function, assumed to have a Gaussian form, centred at $L_i$, and with standard deviation given by the measurement uncertainty $\sigma_i$. In other words, the probability $P(L_{i}$) that an object has a luminosity $L_i$, given the Schechter model, is evaluated  by  integrating with respect to $L$ the convolution of the luminosity function with the Gaussian function $N(L|\{L_i,\sigma_i\})$:

\begin{equation}
\begin{split}
&P(L_{i}) = \frac{ \displaystyle \int \limits _{L_{lim}} ^\infty \Psi(L_{i}) \cdot V_{\mathrm{eff}}(L_{i}) \cdot N(L|\{L_i, \sigma_i\}) \mathrm{d} L}{  \displaystyle \int \limits _{L_{lim}} ^\infty \Psi(L) \cdot V_{\mathrm{eff}}(L) \cdot \mathrm{d}L} \\
&\mathrm{with,} \\
&N(L|\{L_i, \sigma_i\}) = \frac{1}{\sqrt{2\pi}\sigma_i} \mathrm{exp} \left[ - \left( \frac{(L - L_{i})^{2}}{2 \sigma_{i}^{2}}\right)\right]
\end{split}
\end{equation}
\noindent where $L$ is the log luminosity, $V_{\mathrm{eff}}(L)$ is the effective volume from Equation~\ref{eqn:veff} -- which also accounts for the completeness and area coverage of the survey. In the limit where the uncertainties are very small, the Gaussian becomes a delta function and the probability approaches the value defined in the original MLE, thus, recovering the original expression.

In order to test the performance of our modified MLE, we performed a set of simulations (described in Appendix~\ref{apnd:sim_1d}) to reproduce single-line luminosity functions. When the sample includes a small number of bright sources with significant measurement uncertainties the original MLE is less robust than the modified MLE, which marginalizes the probabilities over the measurement uncertainties. For a more detailed discussion, see Appendix~\ref{apnd:sim_1d}. 

\subsection{Setting up the Bivariate LLF}
\label{subsec:setup_blf}

The modified MLE method described in the previous Sections can be extended to the bivariate LLF by replacing the single line LF with the bivariate LLF from Equation~\ref{eqn:BLF} and marginalizing over both the \ha\ and \oiii\  luminosities. The probability for a galaxy with \ha\ log-luminosity  ($x$, in the equations below) and \oiii\ log-luminosity ($y$, in the equations below) can then be written as:

\begin{equation} \label{eqn:bivar_prob}
\begin{split}
P&(x_i,y_i)= \\
&\frac{
\begin{split}
\displaystyle \int\limits_{L_{lim}(z)}^\infty \Psi(x,y;&\vec{\textbf{P}})  \cdot N(x,y|\{x_i,\sigma_{x,i}\},\{y_i,\sigma_{y,i}\}) \cdot \\
& \cdot \frac{dV_{comov}}{dz \cdot d\Omega}(z) \cdot C(x_i,z_i) \cdot \Omega \cdot
\mathrm{d}x \ \mathrm{d}y \ \mathrm{d}z
\end{split}}
{ \displaystyle \int\limits_{L_{lim}(z)}^\infty \Psi(x,y;\vec{\textbf{P}}) \cdot \frac{dV_{comov}}{dz \cdot d\Omega}(z) \cdot C(x,z) \cdot \Omega \cdot \mathrm{d}x \ \mathrm{d}y \ \mathrm{d}z} \\
& \mathrm{with,} \\
&N(x,y|\{x_i,\sigma_{x,i}\},\{y_i,\sigma_{y,i}\}) = \\
& \quad \quad \quad \quad
\frac{1}{2\pi\sigma_{x,i}\sigma_{y,i}} \mathrm{exp} \left[ - \left( \frac{(x - x_i)^2}{2 \sigma_{x,i}^2} + \frac{( y - y_i)^2}{2\sigma_{y,i}^2} \right) \right]
\end{split}
\end{equation}
\noindent where $\sigma_x$ and $\sigma_y$ are the measurement uncertainties in \ha\ and \oiii\ log luminosities, $\Psi$ is now the bivariate luminosity function from Equation~\ref{eqn:BLF}, and $C$ is the completeness function (further described in Section~\ref{subsec:comp}).

The probability for each source is calculated according to Equation~\ref{eqn:bivar_prob}. We construct the log likelihood function as in Equation~\ref{eqn:loglike}. The log likelihood function is maximized and the best fit parameters are obtained using \texttt{scipy.optimize.fmin\_l\_bfgs\_b }. Here, all 5 free model parameters $\vec{\textbf{P}} = [\alpha,L^{\star},\beta,r,\sigma_\mathrm{lnL_{H\alpha}}]$ that define the bivariate luminosity function are left free and determined by the maximizing likelihood function.

The \texttt{scipy.optimize.fmin\_l\_bfgs\_b } is a SciPy package that uses a limited-memory Broyden-Fletcher-Goldfarb-Shanno (BFGS) algorithm in order to find the minimum of a function within the parameter space. The BFGS algorithm approximates the iterative Newton's method for finding solutions to functions. The L-BFGS algorithm modifies  BFGS algorithm to reduce the amount of computer memory used and is well suited for optimizing functions with large number of variables. The version of the algorithm implemented here, L-BFGS-B \citep{zhu97}, was written by Ciyou Zhu, Richard Byrd, and Jorge Nocedal\footnote{\texttt{http://www.ece.northwestern.edu/$\sim$nocedal/lbfgsb.html}}.

Once the best fit parameters, $[\alpha, L^{\star}, \beta, r, \sigma_\mathrm{lnL_{H\alpha}}]$, are obtained, the normalization factor $\phi^\star$ can be computed as:

\begin{equation}
\phi^{\star} = \frac{N}{ \displaystyle \int \limits_{L_{lim}(z)} \Psi(x,y;\vec{\textbf{P}}) \cdot \frac{dV_{comov}}{dz \cdot d\Omega}(z) \cdot C(x,z) \cdot \Omega \mathrm{d}x \ \mathrm{d}y \ \mathrm{d}z}
\end{equation}
where the integration limit is taken to be the median flux limit for all fields, $N$ is the number of sources detected in the survey, $x$ and $y$ are the \ha\ and \oiii\ log luminosities respectively, $\Psi$ is the bivariate luminosity function, $\Omega$ is the solid angle surveyed, ${dV_{comov}}/{dz \ d\Omega}(z)$ is the differential comoving volume at redshift $z$, and $C$ is the completeness function.

We obtain accurate errors for our best-fit model parameters by performing a Markov Chain Monte Carlo (MCMC) analysis using the publicly available \texttt{emcee} Python package\footnote{\texttt{http://dan.iel.fm/emcee/}} \citep{emcee}. We use uninformative uniform priors for our parameters and the likelihood function is defined as Equation~\ref{eqn:loglike} with individual probabilities described by Equation~\ref{eqn:bivar_prob}.

\subsection{Survey Incompleteness}
\label{subsec:comp}
\cite{colbert13} performed an extensive completeness simulation to quantify the survey incompleteness for WISP, which we adopt here. Their completeness is provided as a function of equivalent width and signal-to-noise ratio. In order to implement the completeness function into our formulation, we have to re-parametrize it as a function of flux. The completeness for our sample is given by the \ha\ luminosity (converted to signal-to-noise ratio using the survey limit) and marginalized over the equivalent width distribution for WISP sources, assuming that the equivalent width distribution is independent of the flux.

\begin{equation}\label{eqn:comp}
\displaystyle
C(L,z) = \int C \left( \mathrm{EW}, \mathrm{S/N}=\frac{L}{f_{lim} \cdot 4 \pi d_L^2(z)} \right) \ \mathrm{d}(\mathrm{EW})
\end{equation}

\noindent where $L$ is the line luminosity, EW and S/N are the equivalent width and signal-to-noise of the line, $f_{lim}$ is the flux limit, and $d_L(z)$ is the luminosity distance at redshift $z$.

\section{Fitting the Bivariate LLF at \lowercase{$z\sim1$}}
\label{sec:blf}
\label{subsec:data_blf}

In this section, we use WISP galaxies to derive the best fit parameters of the bivariate LLF at $z\sim 1$, where both \ha\ and \oiii\ emission lines can be detected in the wavelength range covered by the G102$+$G141 spectroscopy. We select the sample to include only galaxies in the redshift range $0.8<z<1.2$ to allow for sufficient sample size while minimizing the impact of an evolving luminosity function over the redshift range. We select 487 galaxies from the 52 WISP fields that satisfy the quality flag cut described in Section~\ref{sec:data} and have \ha\ signal-to-noise ($S/N$) ratio $>5$. Of this sample, 166 galaxies have detected \oiii\ with $S/N>2$. After applying a strict cut of $>$5$\sigma$ in \ha, looking for the \oiii\ line is no longer a blind search. This allows us to relax the $S/N$ cut for the \oiii\ line while maintaining a high quality, pristine sample. Moreover, we do properly account for the errors in the line luminosity during the fitting procedure (see Section~\ref{subsec:modified_mle}).

\begin{figure}[!b]
\centering
\includegraphics[width=0.46\textwidth]{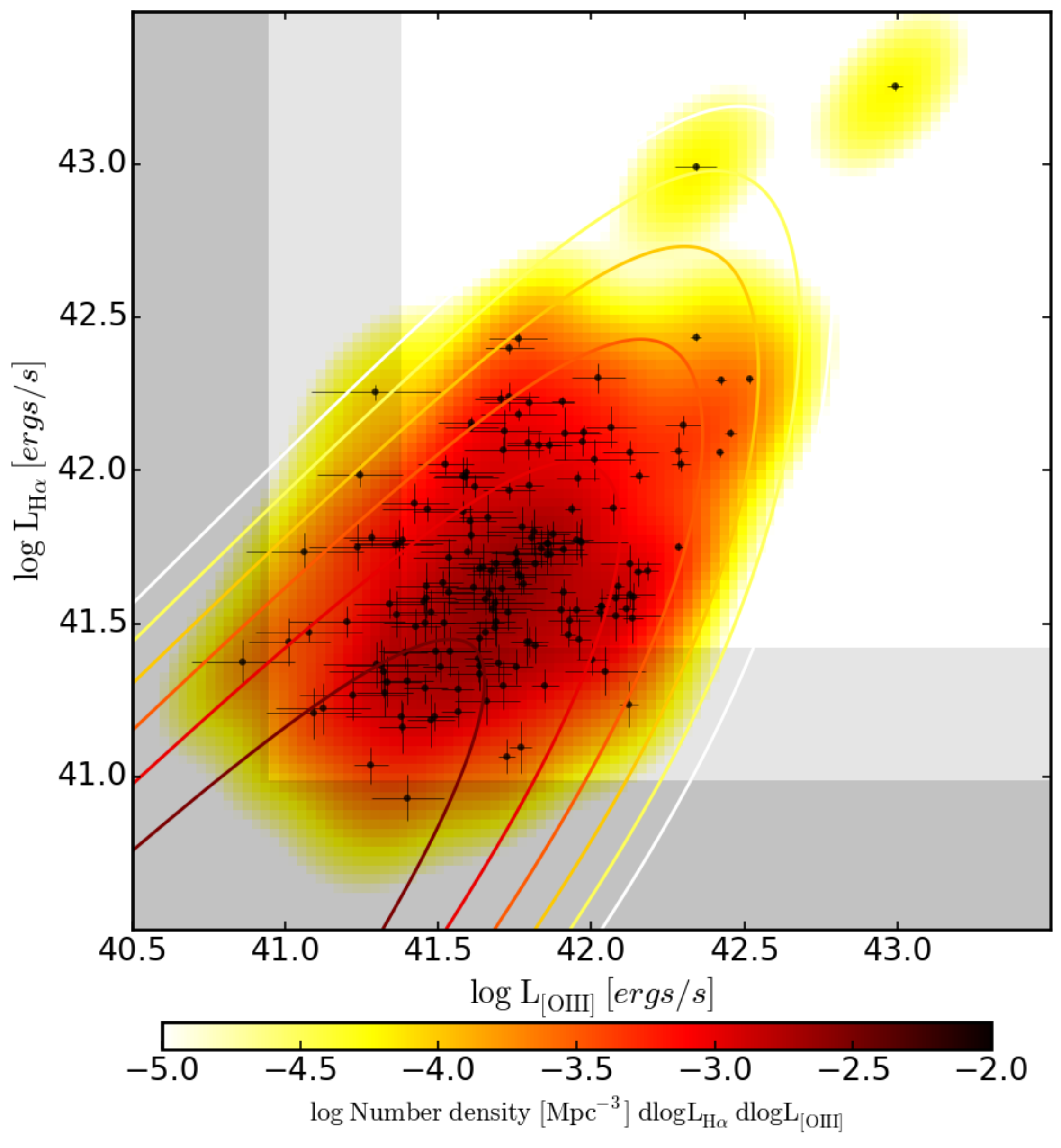}
\caption{WISP data sample for $0.8<z<1.2$ plotted along with the bivariate LLF best fit shown by the contours. The completeness corrected Kernel Density Estimate (KDE) map is plotted in color. The \textit{grey} shaded regions represent our survey limit at $z=0.8$ \textit{(darker)} and $z=1.2$ \textit{(lighter)}.}
\label{fig:fit_blf}
\end{figure}

\begin{figure}[!t]
\centering
\includegraphics[width=0.46\textwidth]{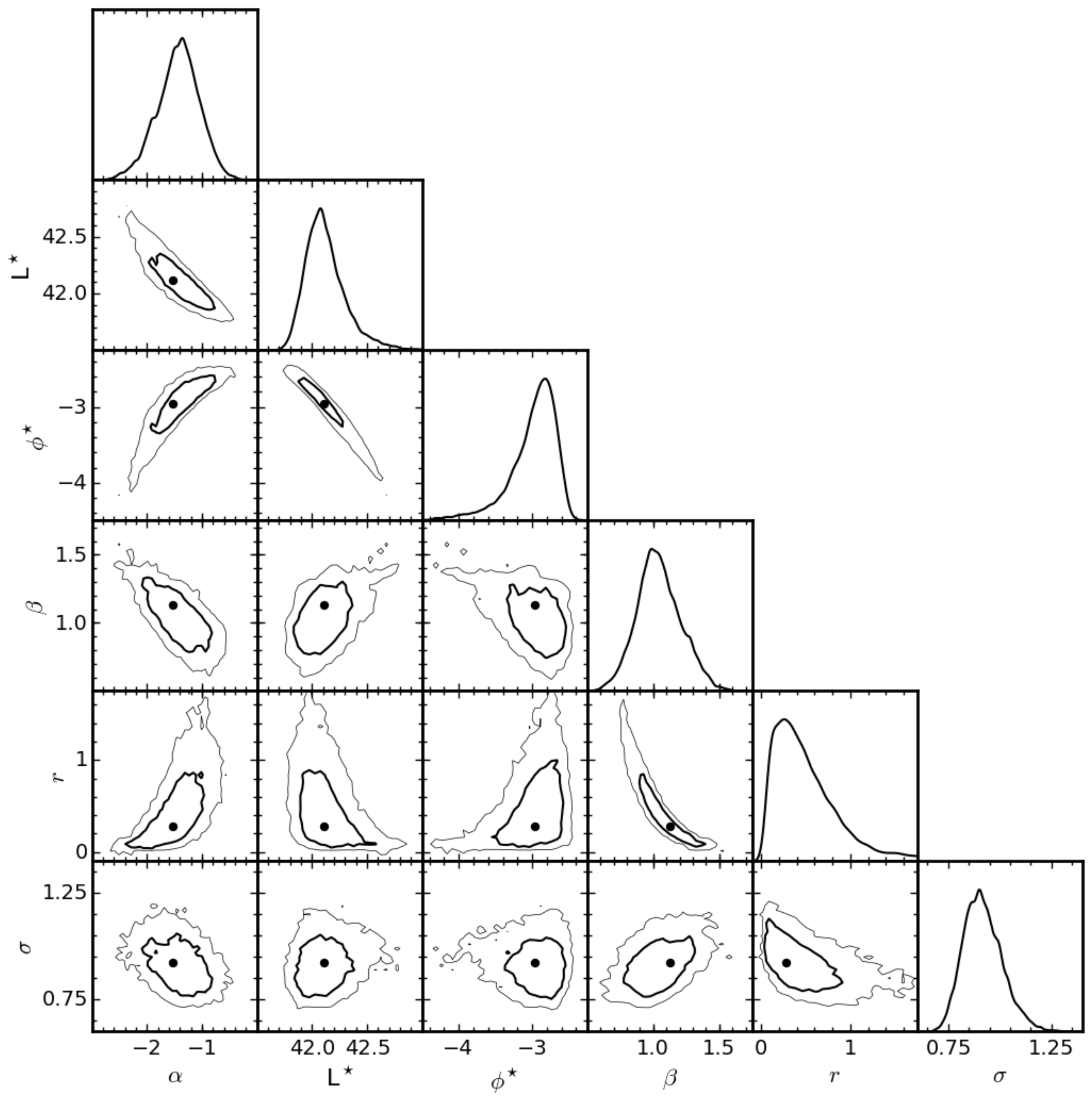}
\caption{The posterior as well as joint-posterior distributions for the bivariate LLF model parameters obtained from MCMC analysis. The best-fit parameters as obtained by the MLE are shown by \textit{black dots}, and for the joint-posterior distributions, the 68\% \textit{(thicker)} and 95\% \textit{(thinner)} confidence contours are shown.}
\label{fig:err_blf}
\end{figure}

\label{subsec:fit_blf}
We fit the \ha --\oiii\ bivariate LLF to the sample of 166 galaxies, following the procedure described in Section~\ref{subsec:setup_blf}. The best-fit model parameters are reported in Table~\ref{tab:blf} and shown in Figure~\ref{fig:fit_blf}. We also perform the MCMC analysis for the model parameters and their posterior distributions are shown in Figure~\ref{fig:err_blf}.

In Figure~\ref{fig:fit_blf}, the black data points show the sample used to fit the bivariate LLF and the contours show the best-fit bivariate LLF. The density map shows the Kernel Density Estimate (KDE) of the data points (detected in both \ha\ and \oiii) corrected for the survey incompleteness. Both the density map and the contours are plotted on the same color-scale. The grey shaded regions show the survey flux limits at $z=0.8$ (darker) and $z=1.2$ (lighter).

\begin{deluxetable}{lccl}
\centering
\tablecolumns{2} 
\tablewidth{0.4\textwidth} 
\tablecaption{Best Fit Parameters for the \ha --\oiii\ bivariate luminosity function fit for $0.8<z<1.2$ sample.}
\tablehead{ 
\colhead{} &\colhead{Parameter} &\colhead{Best-Fit Value} & \colhead{}}
\startdata
&$\alpha$ & -1.5 $^{+0.5}_{-0.2}$ &\\
&log L$^\star$ & 42.1 $^{+0.1}_{-0.2}$ &\\
&log $\phi^\star$ & -2.95 $^{+0.33}_{-0.18}$ &\\
&$\beta$ & 1.13 $^{+0.06}_{-0.26}$ &\\
&$r$ &  0.28 $^{+0.35}_{-0.20}$ &\\
&$\sigma_{\ln \mathrm{L_{H\alpha}}}$\tablenotemark{a} & 0.92 $^{+0.08}_{-0.11}$ &
\enddata
\tablenotetext{a}{$\sigma_{\ln \mathrm{H\alpha}}=\sigma_{\log_{10} \mathrm{H\alpha}}\times \ln(10)$, where $\sigma_{\log_{10} \mathrm{H\alpha}}$ is in dex}
\label{tab:blf}
\end{deluxetable}

\begin{figure}[!b]
\includegraphics[width=0.48\textwidth]{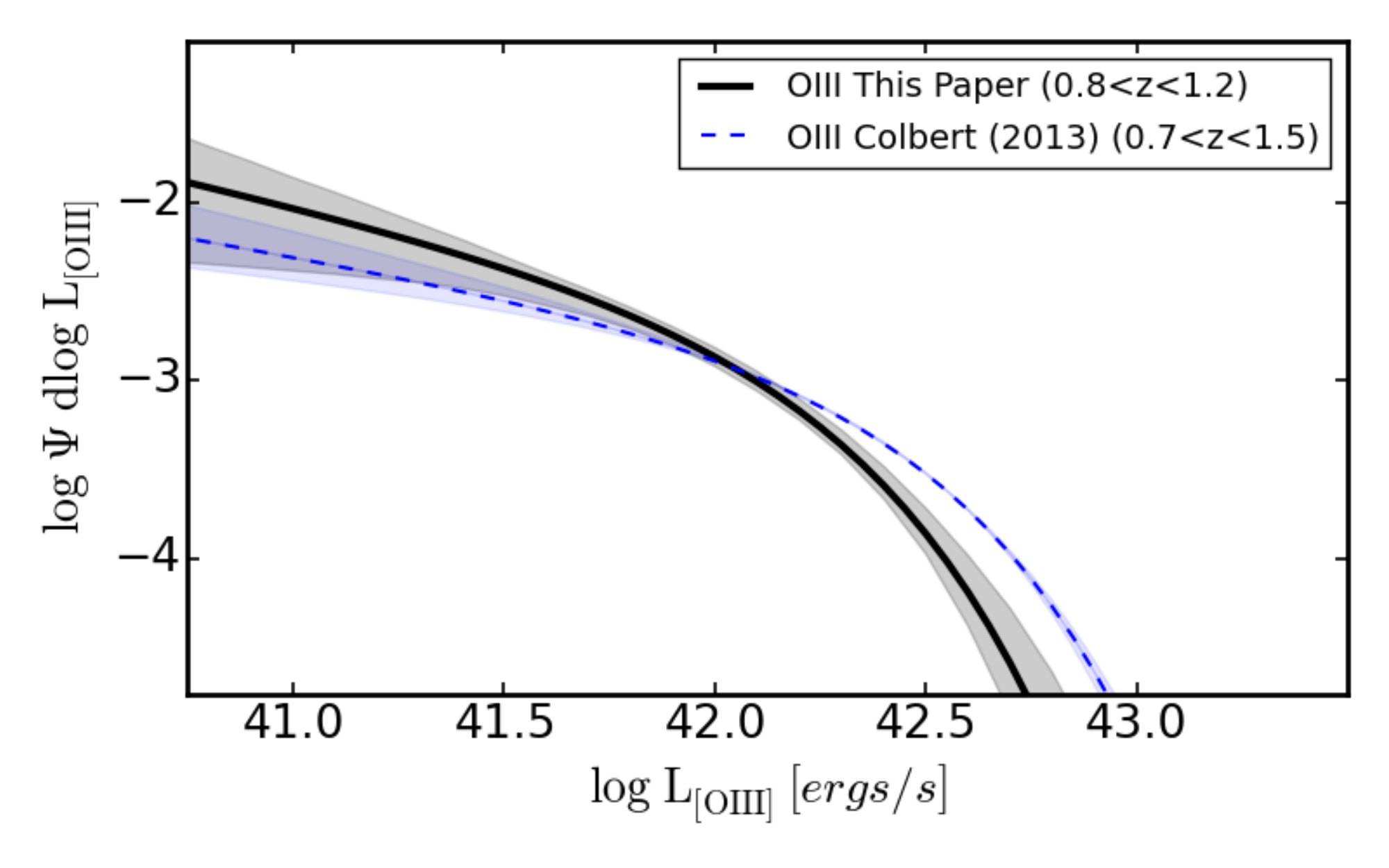}
\caption{The collapsed \oiii\ LLF derived from the best-fit bivariate LLF from Figure~\ref{fig:fit_blf} compared with the \oiii\ luminosity function from \cite{colbert13}. The shaded regions represent the $1\sigma$ deviations in the best-fit parameters.}
\label{fig:fit_oiii_blf}
\end{figure}

With the best fit parameters for the bivariate LLF, we can derive the single-line luminosity function by marginalizing over the nuisance dimension (e.g., the \oiii\ luminosity function can be computed by integrating over the \ha\ nuisance dimension). In Figure~\ref{fig:fit_oiii_blf}, we compare the marginalized \oiii\ luminosity function with the result of the LF of \cite{colbert13}, computed for galaxies in the $0.7<z<1.5$ redshift range. Our new estimate of the marginalized \oiii\ LF has a faint end slope consistent within the errors with the slope derived by \citet[][$-1.5^{+0.5}_{-0.2}$ versus $-1.4\pm0.15$]{colbert13}. The characteristic $L_{\mathrm{[OIII]}}^\star$ luminosity is somewhat lower, although the results are within 2$\sigma$ from each other ($42.1^{+0.1}_{-0.2}$ versus $42.34 \pm 0.06$). However, we note that the two analyses are not expected to provide the same best-fit parameters for various reasons: first the narrower redshift range used in our work minimizes the effect of the evolution of $L^\star$ over the redshift interval, and  second our fitting method is not affected by the arbitrary choice of bin size and centers, as well as it accounts for uncertainties in the line luminosities.  

Now, we can compute the \ha\ LF by integrating the bivariate LLF over the \oiii\ dimension and accounting for the \oiii\ non-detection rate. For fitting the bivariate LLF, we only used sources detected in both \ha\ and \oiii. However, there is a significant fraction of sources that are detected in \ha\ but not in \oiii\ despite being within the wavelength coverage, due to the significant intrinsic scatter in the \ha --\oiii\ relation as well as the sensitivity limits of our survey. Thus, the bivariate LLF parameters obtained above reproduce the number density of galaxies detected in both lines, but  underestimates the number density of \ha-emitters selected regardless of their \oiii\ luminosity. With the goal of obtaining \ha\ number counts from the \oiii\ luminosity function, we compute a statistical correction that accounts for the fraction of \ha-emitters missed due to \oiii\ non-detections as a function of the \ha\ luminosity. The non-detection correction term is then applied when collapsing the bivariate LLF to obtain the single-line \ha\ LF. 

\begin{figure}[!t]
\includegraphics[width=0.48\textwidth]{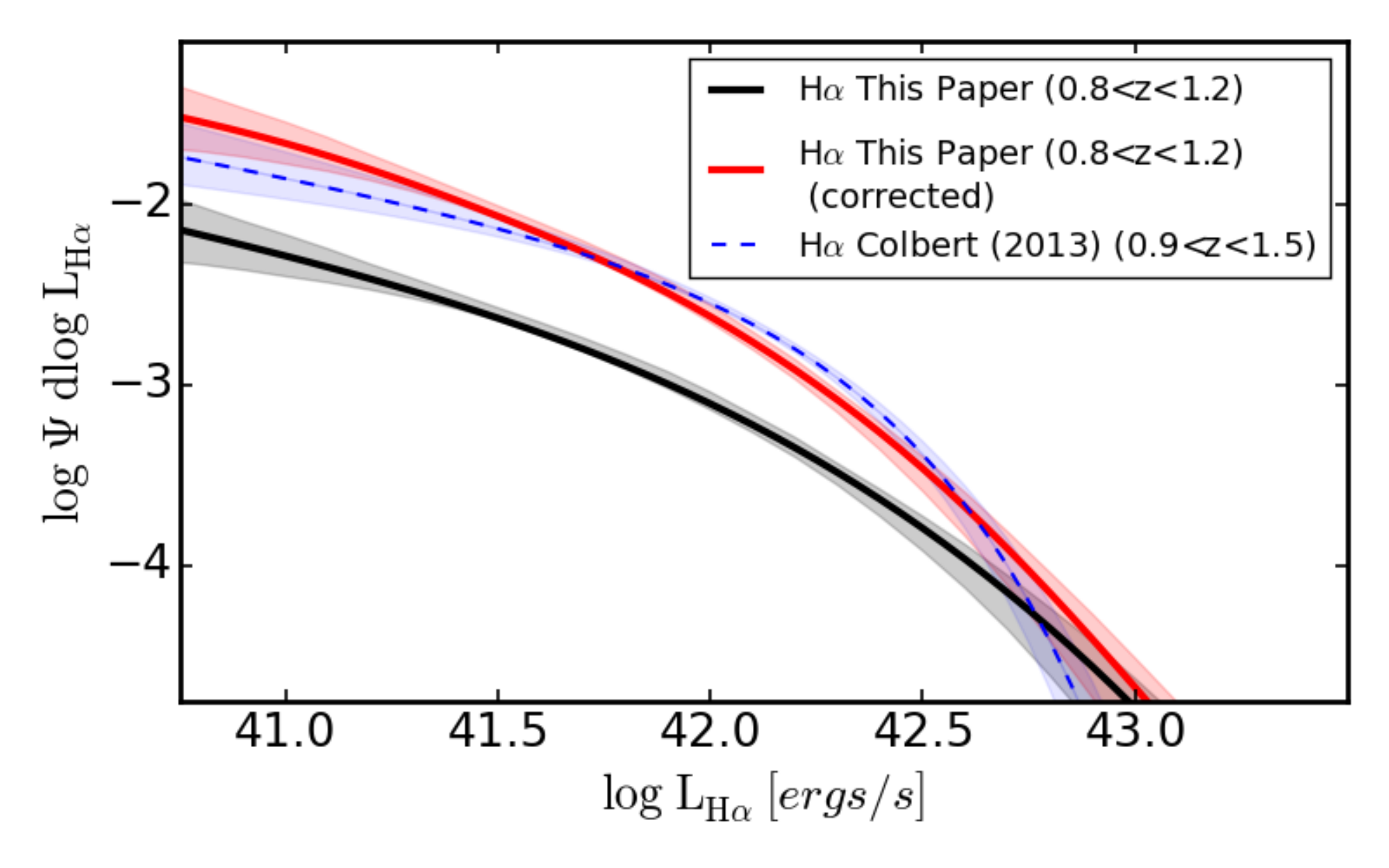}
\caption{The collapsed \ha\ LLF derived from the best-fit bivariate LLF from Figure~\ref{fig:fit_blf} compared with the \ha\ luminosity function from \cite{colbert13}. The uncorrected (in \textit{black}) and non-detection corrected (in \textit{red}) LFs are shown. The shaded regions represent the $1\sigma$ deviations in the best-fit parameters.}
\label{fig:fit_ha_blf}
\end{figure}

We compute the non-detection correction term as the fraction of galaxies below the detection limit in \oiii, in bins of \ha\  luminosity. For this analysis we used all galaxies in the redshift range $0.7<z<1.5$, where both emission lines are covered in the spectroscopic observations.  The non-detection correction ranges between 100\% and $\sim$400\%, for \ha\ luminosities between $10^{43}$ and $10^{41}$ ergs s$^{-1}$, respectively.  In Figure~\ref{fig:fit_ha_blf}, we show the marginalized \ha\ LF without the non-detection correction (black solid line) and with the correction applied (red line), and compare the results with the \ha\ LF derived in Colbert et al. (2013). The marginalized \ha\ LF corrected for the \oiii\ non-detection fraction is excellent agreement with the Colbert et al. (2013) \ha\ LF. We note that in our formulation, the \ha\ LF marginalized function over the \oiii\ dimension is not exactly a Schechter function, but as Figure~\ref{fig:fit_ha_blf} shows, it is very close to it.

\section{Estimating \ha\ from \oiii\ at \lowercase{$z\sim1$}}
\label{sec:z1}
The main goal of this work is  to estimate the \ha\ number counts at $z\sim2$, starting from the \oiii\ LF  at the same redshift.  Here, we test how accurately the \ha\ number counts  obtained with this approximation reproduce the known \ha\  number counts at $z\sim1$, obtained through direct integration of the \ha\ LF. To this aim we use the sample of 129  \oiii-selected galaxies in the $0.8<z<1.2$ redshift range  and with \oiii\ $S/N>5$, along with the same quality flag cuts as described in Section~\ref{sec:data}. The \oiii\ single line luminosity function is computed using the modified MLE and the completeness analysis from \cite{colbert13}. The best-fit parameters for the \oiii\ LF at $z\sim1$ are reported in Table~\ref{tab:z1}. 

\begin{figure}[!b]
\includegraphics[width=0.48\textwidth]{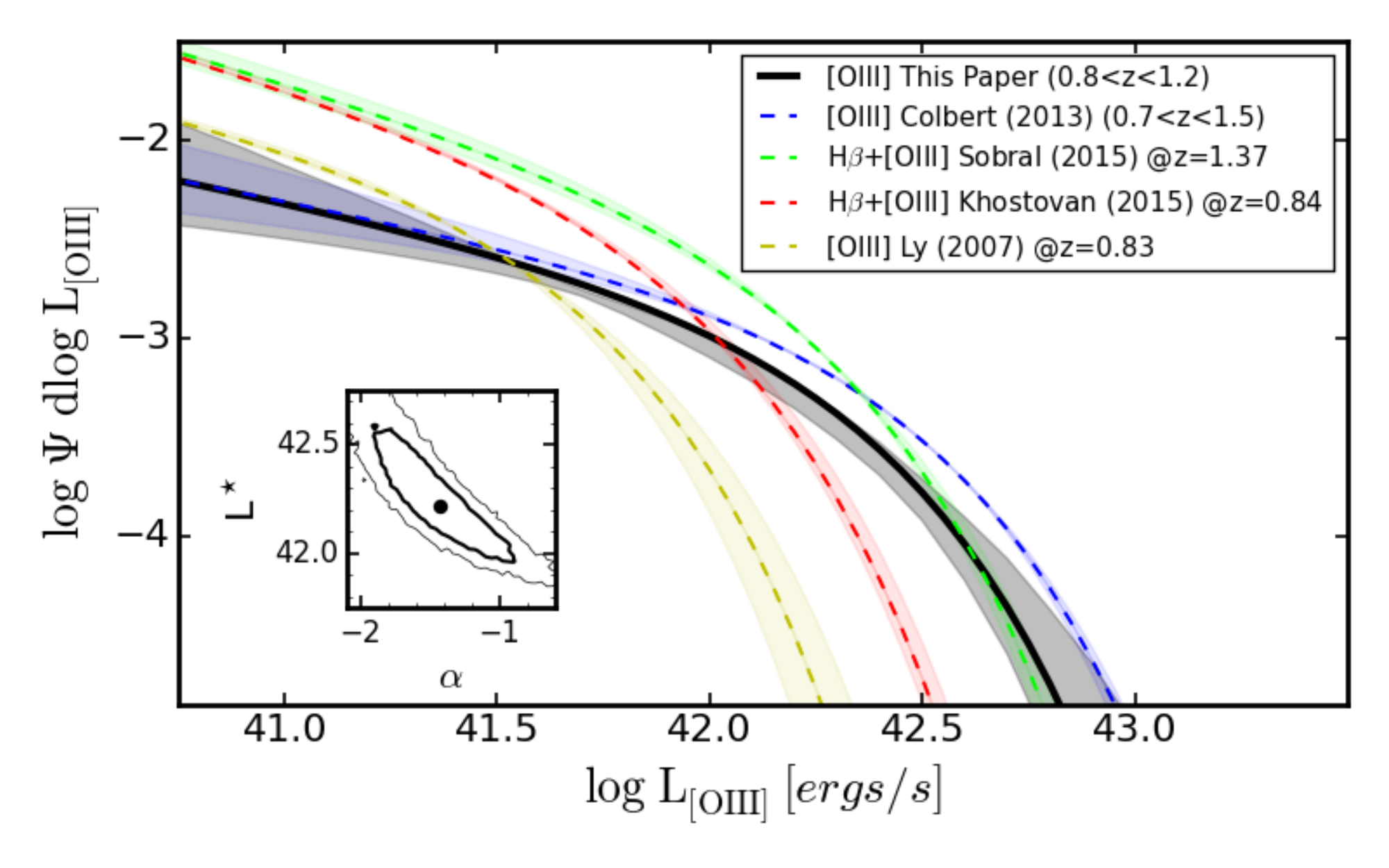}
\caption{The best-fit \oiii\ LLF derived for $0.8<z<1.2$ WISP dataset compared with the \oiii\ luminosity function from \cite{colbert13} as well as other estimates from the literature. The shaded regions represent the $1\sigma$ deviations in the best-fit parameters. The inset shows the joint-posterior distribution of $\alpha$ and $L^\star$ from MCMC analysis for our best-fit \oiii\ LF.}
\label{fig:fit_oiii_z1}
\end{figure}

Figure~\ref{fig:fit_oiii_z1} shows our best-fit \oiii\ single line luminosity function as well as the results from the literature. As noted before, we find a good agreement between our and \citet{colbert13} measurements of the LF on the WISP datasets. The rise in the faint-end for \cite{khostovan15} and \cite{sobral15} \hb +\oiii\ LFs can be attributed to the \hb\ emitters in their sample. However, we note that the variation among different measurements is still substantial, especially at the bright end -- the number density of $L^\star$ galaxies (i.e., $\sim10^{42}$ ergs s$^{-1}$) varies by almost an order of magnitude. This comparison clearly shows that the nominal errors typically quoted on the best-fit LF parameters generally do not provide an adequate measurement of the actual variation observed in LF determination. The source of this variation is most likely systematic (e.g., different selection techniques provide systematically brighter/fainter samples, galaxy clustering could systematically enhance/suppress number counts in small-area fields, and so on), and need to be accounted for in predictions used to optimize large galaxy redshift surveys aiming to constrain dark energy. 

\begin{deluxetable}{lccl}
\centering
\tablecolumns{2} 
\tablewidth{0.4\textwidth} 
\tablecaption{Best Fit Parameters for the \oiii\ exclusive luminosity function fit for $0.8<z<1.2$ sample.}
\tablehead{ 
\colhead{} &\colhead{Parameter} &\colhead{Best-Fit Value} & \colhead{}}
\startdata
&$\alpha$ &  -1.42 $^{+0.23}_{-0.43}$ &\\
&log L$^\star$ & 42.21 $^{+0.22}_{-0.18}$ &\\
&log $\phi^\star$ &  -3.17 $^{+0.27}_{-0.39}$ &
\enddata
\label{tab:z1}
\end{deluxetable}

\begin{figure}[!t]
\includegraphics[width=0.48\textwidth]{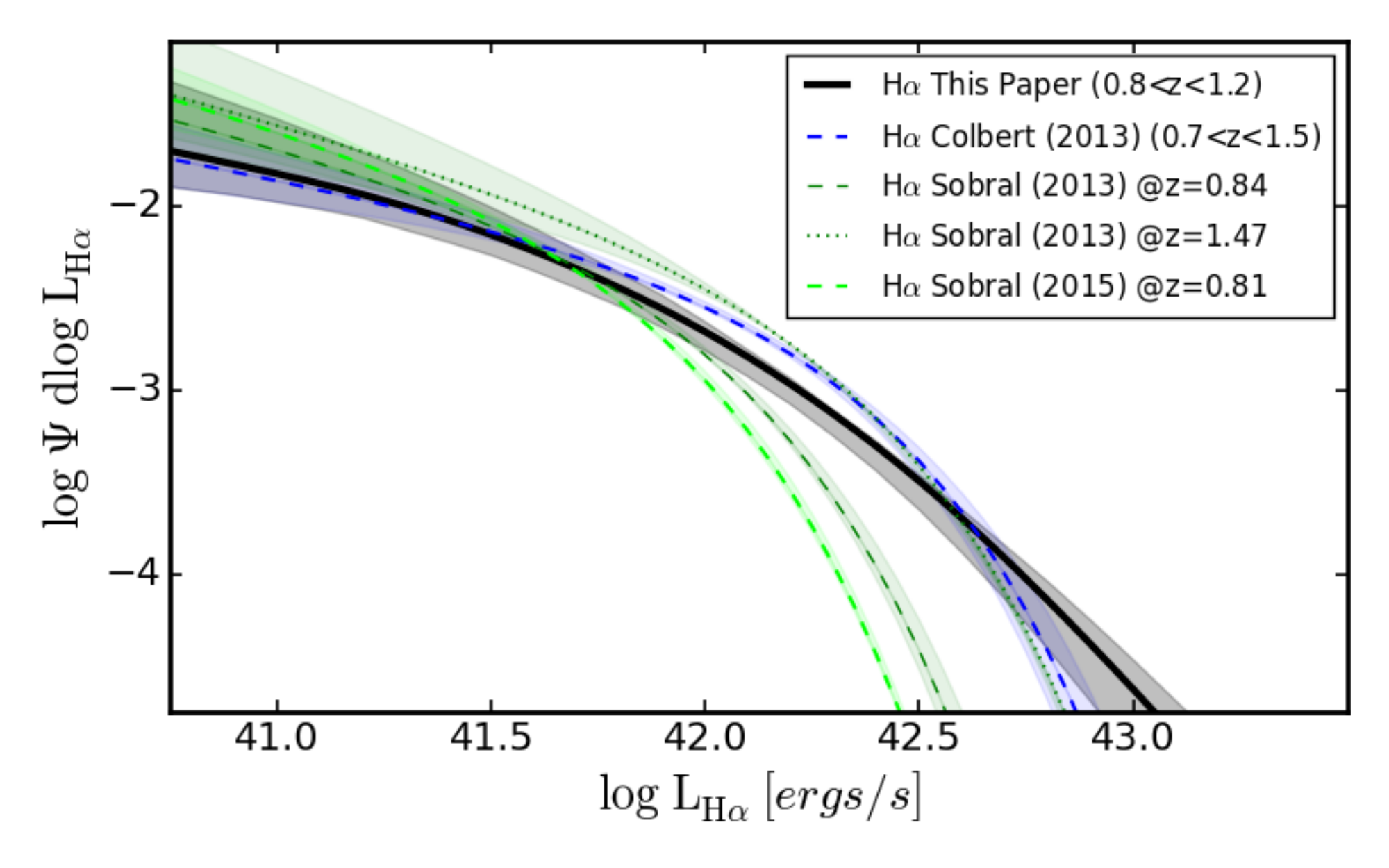}
\includegraphics[width=0.48\textwidth]{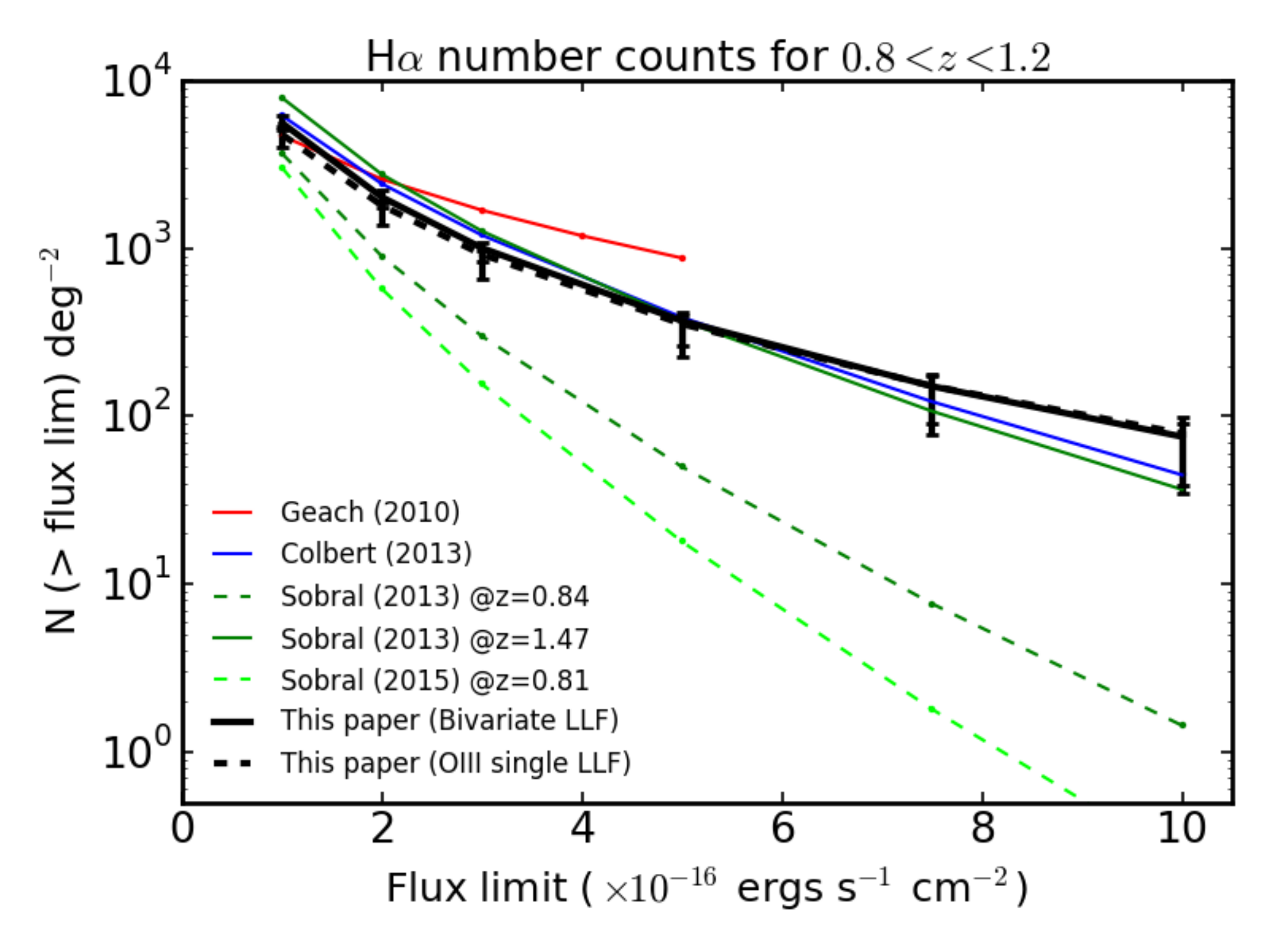}
\caption{\textit{Top:} The collapsed \ha\ LLF  for $0.8<z<1.2$ derived using the best fit \oiii\ LLF in Figure~\ref{fig:fit_oiii_z1} and the best fit bivariate LLF parameters $\beta$, $r$, and $\sigma_{\mathrm{ln H\alpha}}$ from Table~\ref{tab:blf}. \textit{Bottom:} The \ha\ number counts estimated for $z\sim1$ from our \ha\ LF, along with estimates from other groups in the literature. The errors on our estimate accounts for the uncertainties in the best-fit parameters in addition to the normal Poisson errors. All number counts are corrected for survey incompleteness.}
\label{fig:fit_ha_z1}
\end{figure}

\label{subsec:fit_ha_z1}
Using the best-fit Schechter parameters obtained for the \oiii\ single line LF, we reconstruct  the bivariate LLF using the parameters $\beta$, $r$, and $\sigma_{\mathrm{ln H\alpha}}$ from Table~\ref{tab:blf}. As done in Section~\ref{sec:blf}, we compute the \ha\ luminosity function by marginalizing over the \oiii\ dimension. In Figure~\ref{fig:fit_ha_z1}, we compare the \ha\ single-line luminosity function obtained from the \oiii\ LF  with the direct estimate from Colbert et al. (2013). Clearly, the two LF agree very well, as demonstrated also by the cumulative number counts shown in  Figure~\ref{fig:fit_ha_z1}, where we show the \ha\ number counts obtained both from the \oiii--only fit and  from the bivariate LLF fit from Section~\ref{sec:blf} (black solid and dashed line, respectively). The errors on the number counts account for the uncertainties in the best fit parameters in addition to the normal Poisson errors. Also note that the number counts are corrected for survey incompleteness. As evident from the figure, the recovered \ha\ number counts from the \oiii-only fit agree extremely well with the bivariate version as well as from direct integration of the \ha\ LF. We add for completeness the number counts from Sobral et al. (2012) and Geach et al. (2010). The differences between the WISP dataset and these two works are discussed in detail in Colbert et al. (2013).

\section{Estimating \ha\ from \oiii\ at \lowercase{$z\sim2$}}
\label{sec:z2}
Having demonstrated the feasibility of our procedure at $z\sim 1$, we now apply it to the redshift 2 case.
In Section~\ref{sec:ha_oiii_trend}, we compared the \ha --\oiii\ correlation for SDSS ($0.2<z<0.3$) and WISP ($0.8<z<1.2$) data. There is little evidence for significant evolution of the \ha --\oiii\ correlation between $z\sim0.25$ and $z\sim1$. Continuing with the assumption that the \ha --\oiii\ correlation from $z\sim1$ also holds at $z\sim2$, we use the sample of WISP \oiii--emitters at $z\sim2$ together with the \ha --\oiii\ ratio parameters obtained at $z\sim1$ (see Section~\ref{subsec:fit_blf}) to derive the $z\sim2$ \ha\ number counts. We follow the same steps as in Section~\ref{sec:z1}. Namely, we first fit a Schechter model to the \oiii-only line LF. Next, we use the best-fit parameters together with $r$, $\sigma_{L_{lnH\alpha}}$, and $\beta$ from Table~\ref{tab:blf} to construct the $z\sim2$ bivariate LLF. Finally, we compute the marginalized $z\sim2$ \ha\ LF, and integrate it to obtain the \ha\ number counts.

\begin{figure}[!b]
\includegraphics[width=0.48\textwidth]{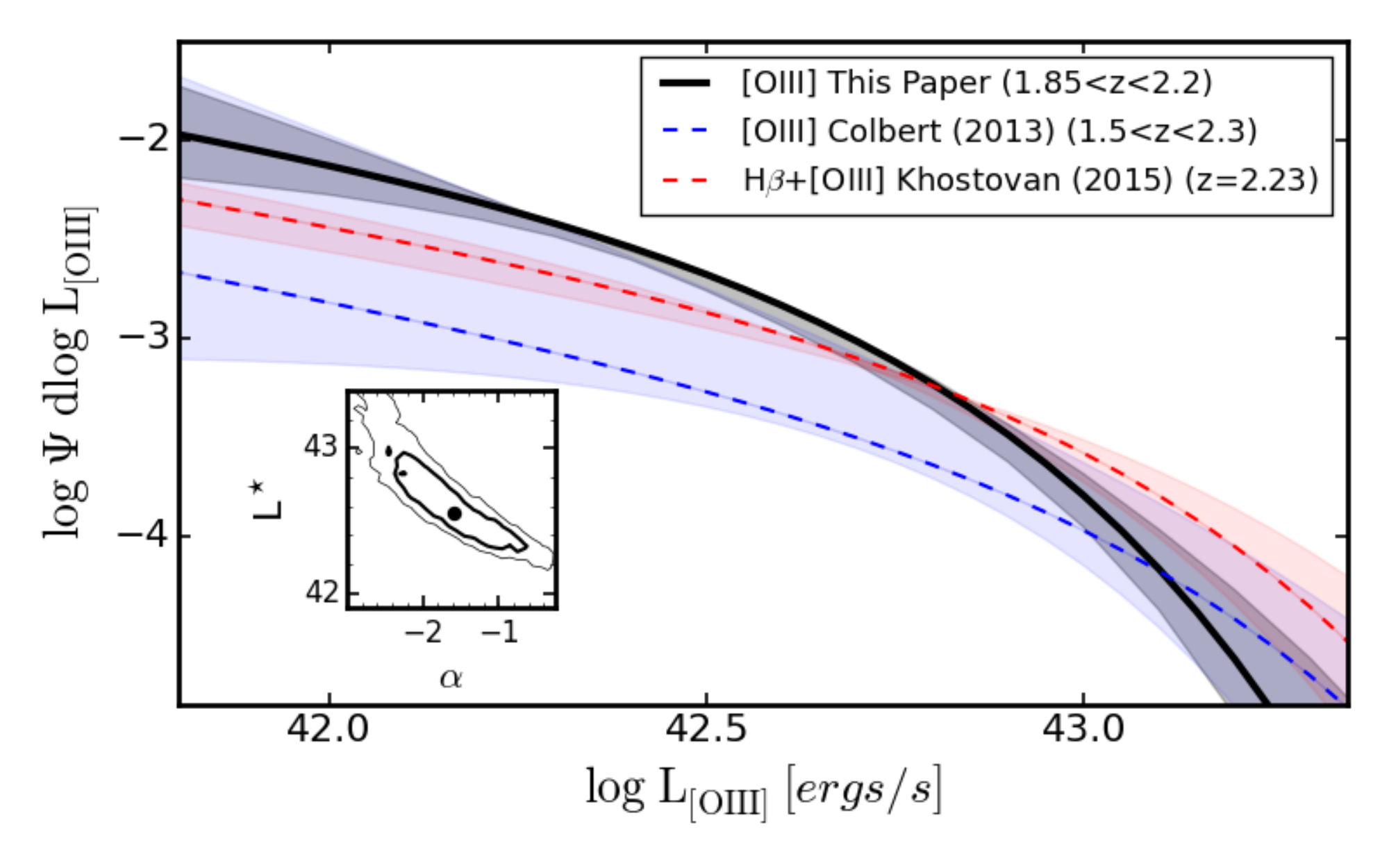}
\caption{The best-fit \oiii\ LLF at $z\sim2$ derived from WISP data plotted along with other estimates from the literature. The shaded regions represent the $1\sigma$ deviations in the best-fit parameters. The inset shows the joint-posterior distribution of $\alpha$ and $L^\star$ from MCMC analysis for our best-fit \oiii\ LF.}
\label{fig:fit_oiii_z2}
\end{figure}

The $z\sim2$ sample consists of 91 WISP \oiii-emitting galaxies selected to be in the redshift range $1.85<z<2.2$, to have \oiii\  $S/N>5$, and redshift quality flags $<16$. The \oiii\ single line luminosity function is fit using the modified MLE, accounting for the measurement uncertainties, and the completeness analysis from \cite{colbert13}. An additional completeness factor is applied in order to account for the loss of high-$z$ \oiii\ emission lines, due to the inability to resolve the doublet, as discussed in \cite{colbert13}.  The best-fit parameters for the \oiii\ LF at $z\sim2$ are reported in Table~\ref{tab:z2}. Our best-fit $z\sim2$ \oiii\ luminosity function is plotted in Figure~\ref{fig:fit_oiii_z2}, along with the result from \cite{colbert13} and other estimates from the literature. Our best fit LF shows a slightly steeper faint-end slope and a higher $\phi^\star$ than what was derived by \cite{colbert13} on a smaller sample. The errors in our previous work, however, were substantial, and the differences are not significant. 

\begin{deluxetable}{lccl}
\centering
\tablecolumns{2} 
\tablewidth{0.4\textwidth} 
\tablecaption{Best Fit Parameters for the \oiii\ exclusive luminosity function fit for $1.85<z<2.2$ sample.}
\tablehead{ 
\colhead{} &\colhead{Parameter} &\colhead{Best-Fit Value} & \colhead{}}
\startdata
&$\alpha$ &  -1.57 $^{+0.28}_{-0.77}$ &\\
&log L$^\star$ & 42.55 $^{+0.28}_{-0.19}$ &\\
&log $\phi^\star$ & -2.69 $^{+0.31}_{-0.51}$ &
\enddata
\label{tab:z2}
\end{deluxetable}

\begin{figure}[!t]
\includegraphics[width=0.48\textwidth]{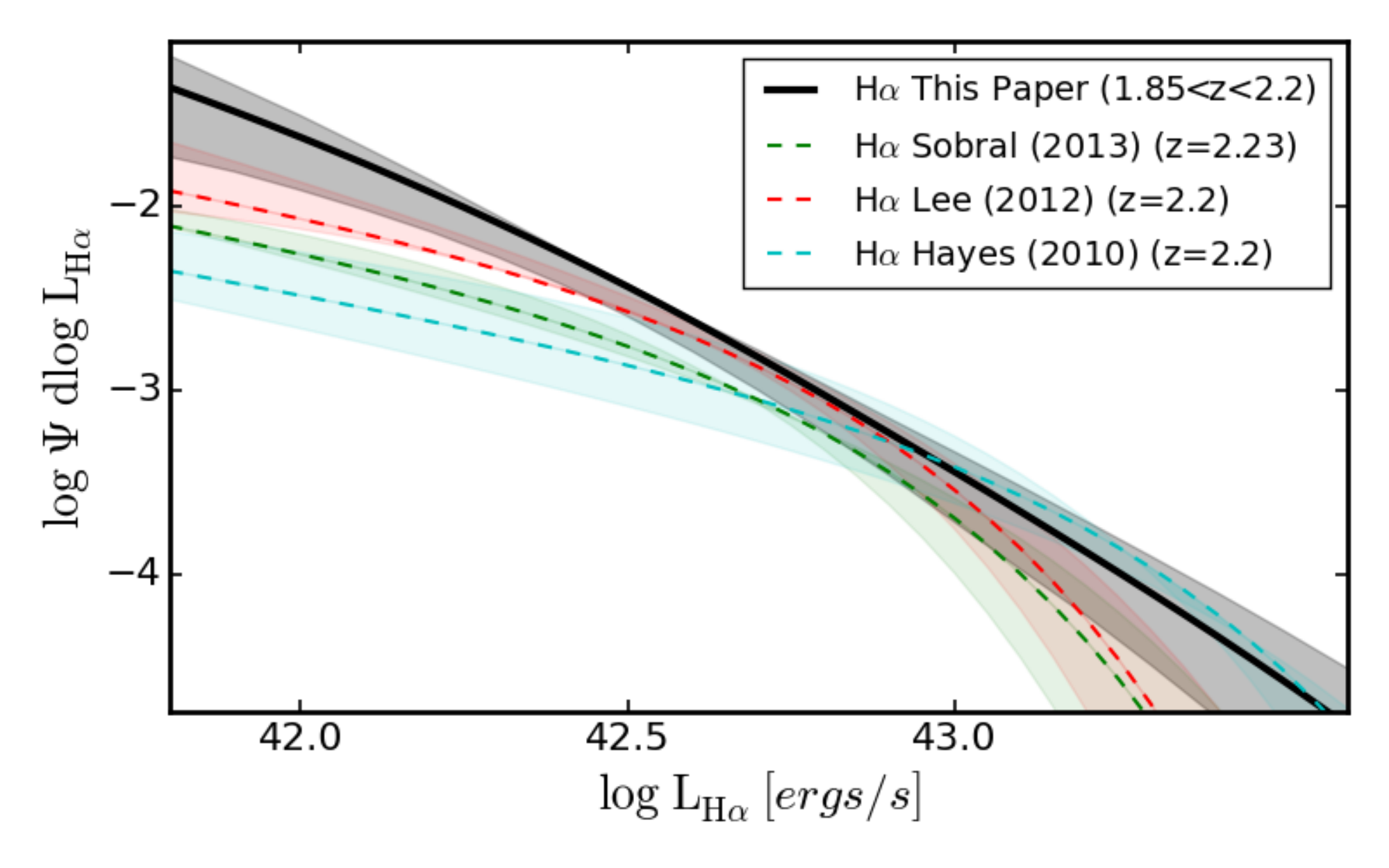}
\includegraphics[width=0.48\textwidth]{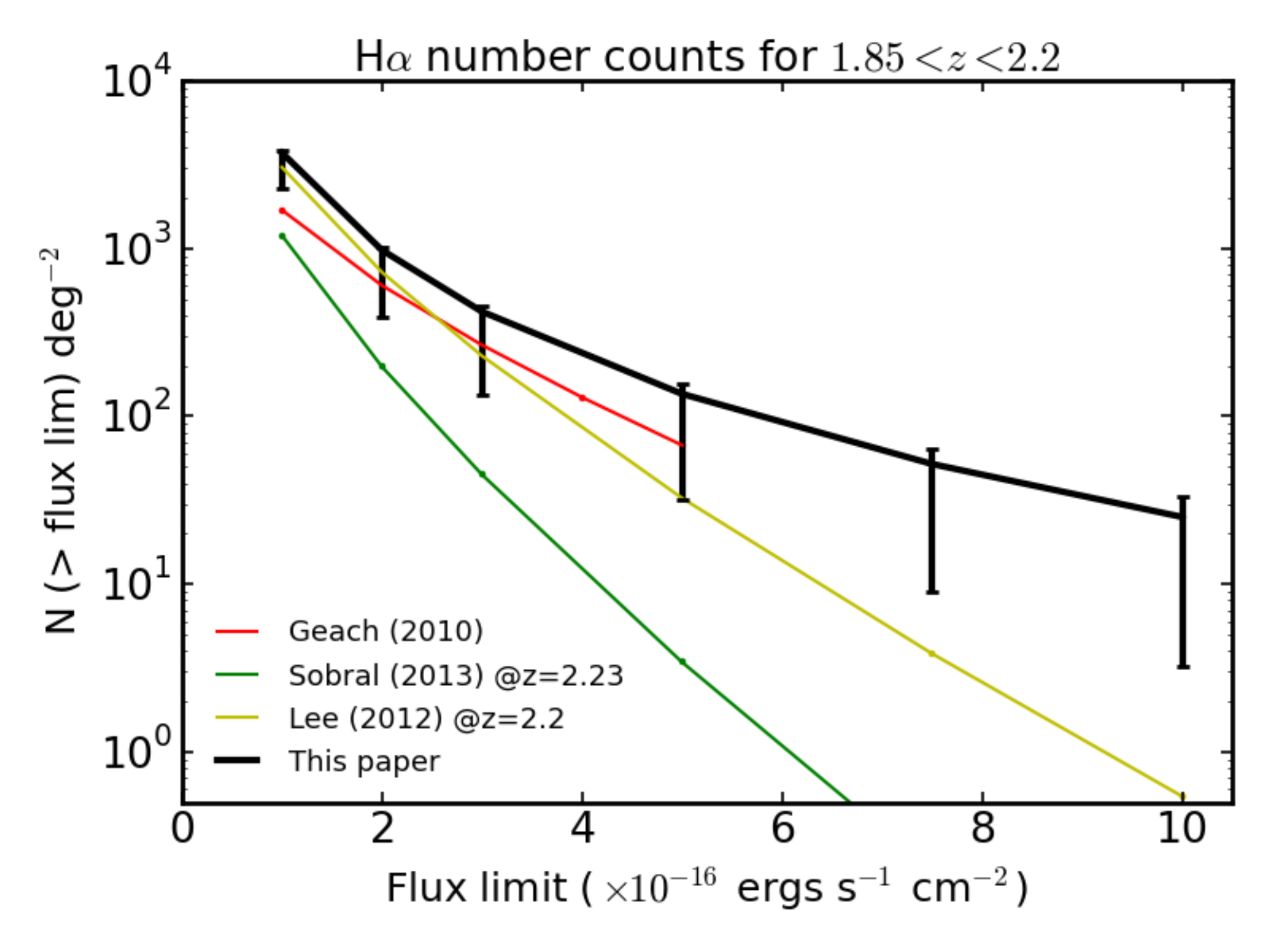}
\caption{\textit{Top:} The collapsed \ha\ LLF at $z\sim2$ derived using the best fit \oiii\ LLF and the best fit bivariate LLF parameters from Table~\ref{tab:blf}. The shaded regions represent the $1\sigma$ deviations in the best-fit parameters. \textit{Bottom:} The \ha\ number counts estimated for $z\sim2$ from our \ha\ LF, along with estimates from other groups in the literature. The errors on our estimate accounts for the uncertainties in the best-fit parameters in addition to the normal Poisson errors. All number counts are corrected for survey incompleteness.}
\label{fig:fit_ha_z2}
\end{figure}

Using the best-fit \oiii\ single line luminosity function, the bivariate LLF is reconstructed using the parameters $\beta$, $r$, and $\sigma_{\mathrm{ln H\alpha}}$ from Table~\ref{tab:blf} and the \ha\ luminosity function is obtained by marginalizing over the \oiii\ dimension. In Figure~\ref{fig:fit_ha_z2}, our best-fit $z\sim2$ \ha\ single line luminosity function obtained from fitting just the \oiii\ data is shown alongside other estimates from the literature. The variation among different determinations is large,  probably because of systematic uncertainties due to different selection techniques, area covered, and procedures used for the estimates of the Schechter parameters. As noted before, these systematic effects are typically not accounted for in the errors quoted alongside the best-fit estimates of the parameters. Thus, the shaded areas shown in Figure~\ref{fig:fit_ha_z2}, are lower limits to the real variation of the volume density at each \ha\ luminosity.

\section{\ha\ Number counts}
We use the collapsed \ha\ luminosity function to compute the $1.85<z<2.2$ \ha\ number counts down to the a range of limiting flux values expected to be reached by future dark energy surveys, and plot the results in Figure~\ref{fig:fit_ha_z2}. For comparison, we have also plotted numbers from \cite{geach10}, \cite{sobral13}, and \cite{lee12}. The numbers for \cite{geach10} are reduced by a factor of ln(10) to account for an error in the published article, resulting from improper conversion of $\Psi(\mathrm{log}L)$ luminosity functions to the standard $\Psi(L)$ luminosity functions.  In Figure~\ref{fig:fit_ha_z2}, our number counts are higher than previous estimates at all flux limits, although below $5 \times 10^{-16}$ ergs s$^{-1}$ cm$^{-2}$, \cite{geach10}, \cite{lee12}, and our work agree within the error bars. The \cite{sobral13} counts are still lower than any previous as well as our estimates. At  brighter fluxes, the variation among the various estimates is very large. Number counts of bright rare galaxies, however, are strongly affected by sample variance. The WISP number counts suffer less from this effect, because of the observing strategy (52 independent fields scattered over the full sky).

\begin{figure}[!b]
\includegraphics[width=0.48\textwidth]{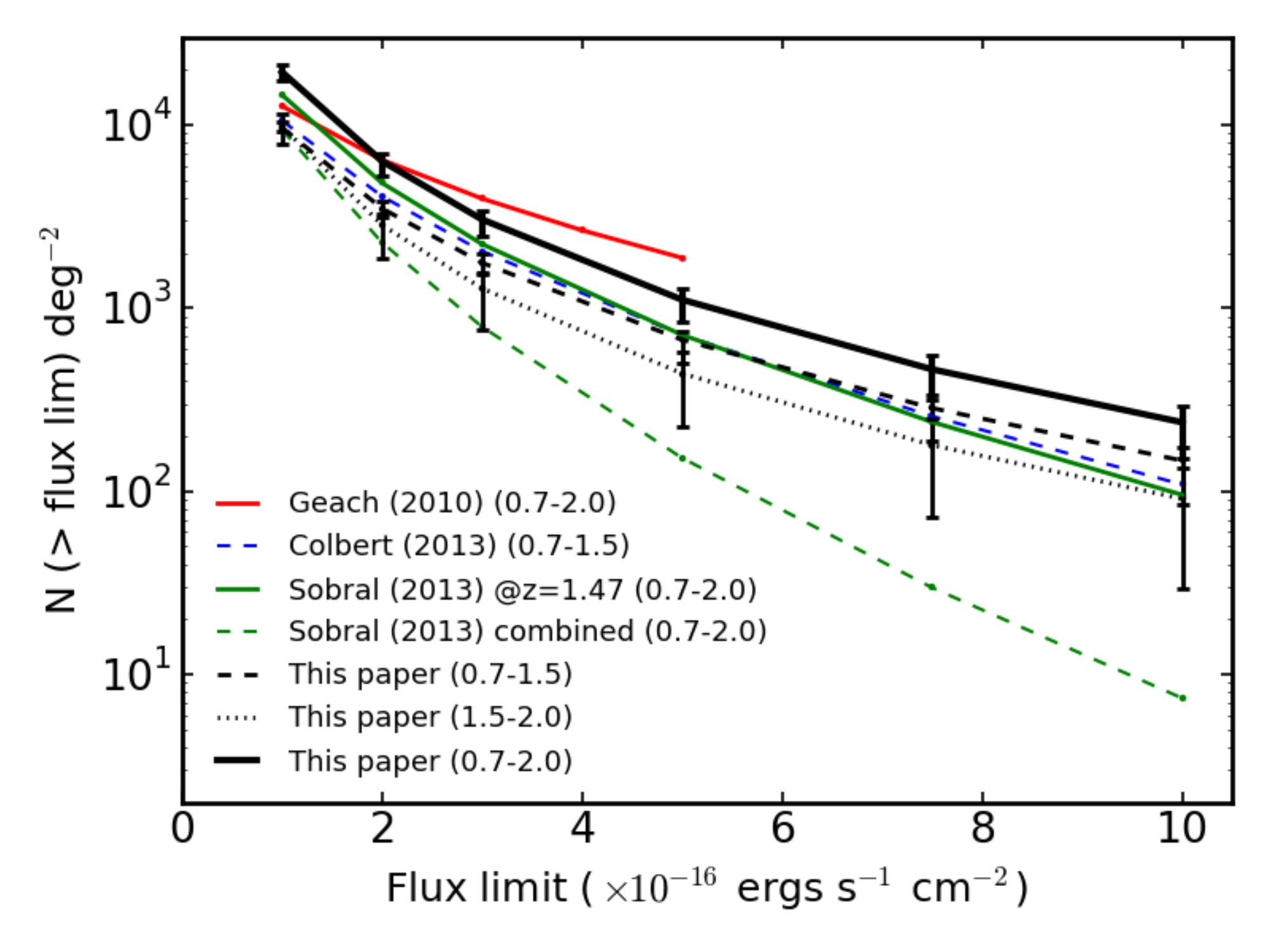}
\caption{The \ha\ number counts estimated for $0.7<z<2.0$ -- relevant redshift range for future surveys to cover \ha. The redshift range for each estimate is reported in parenthesis in the legend. Our estimate \textit{solid black} is split into two ranges: $0.7<z<1.5$ \textit{(dashed black)} and $1.5<z<2$ \textit{(dotted black)}, which use the $z\sim1$ and $z\sim2$ \ha\ LFs, respectively. We also plot two estimates from \cite{sobral13}: \textit{solid green} line represents the \ha\ number counts estimated using the \ha\ LF derived at $z=1.47$ and \textit{dashed green} line represents the sum of number counts estimated over the redshift ranges 0.7--1.2, 1.2--1.85, 1.85--2.0 using \ha\ LFs derived at $z$=0.84, 1.47, 2.23, respectively. The errors on our estimate accounts for the uncertainties in the best-fit parameters in addition to the normal Poisson errors. All number counts are corrected for survey incompleteness.}
\label{fig:counts_tot}
\end{figure}

Finally, we also provide number counts for the whole redshift range expected to be covered by the upcoming dark energy surveys. Figure~\ref{fig:counts_tot} shows the expected \ha\ number counts as a function of survey flux limit for the redshift range $0.7<z<2$. The redshift range is broken into two: $0.7<z<1.5$, where the best-fit \ha\ LF from Section~\ref{sec:blf} is used, and $1.5<z<2$, where the result from Section~\ref{sec:z2} is used to compute the number counts. For comparison, we again plot the \cite{colbert13} (for $0.7<z<1.5$) and \cite{geach10} (for $0.7<z<2.0$) number count estimates. For the redshift range $0.7<z<2$, we expect $\sim$3000 galaxies/deg$^2$ for a flux limit of $3 \times 10^{-16}$ ergs s$^{-1}$ cm$^{-2}$ (the proposed depth of \textit{Euclid} galaxy redshift survey, see \cite{laureijs11}) and $\sim$20,000 galaxies/deg$^2$ for a flux limit of $\sim 10^{-16}$ ergs s$^{-1}$ cm$^{-2}$ (the baseline depth of \textit{WFIRST} galaxy redshift survey, see \cite{spergel15}), when probing with \ha. Number counts for various redshift ranges and limiting fluxes are summarized in Table~\ref{tab:counts}. These number counts have been corrected for survey incompleteness as well as for \nii\ contamination as discussed in Section~\ref{sec:data}. 

The planned spectral resolution for the \textit{Euclid} mission, at the time of writing, is R$\sim$250 \citep{laureijs11}, which will not be able to resolve \ha+\nii. Hence, we also provide the numbers counts that are not corrected for the \nii\ contamination -- these are summarized in Table~\ref{tab:counts_n2}. For the redshift range $0.7<z<2$, we expect $\sim$5700 galaxies/deg$^2$ for the \ha+\nii\ flux limit of $3 \times 10^{-16}$ ergs s$^{-1}$ cm$^{-2}$.

\begin{deluxetable*}{cccccc}
\centering
\tablecolumns{6} 
\tablewidth{\textwidth} 
\tablecaption{Cumulative \ha\ number counts (after applying \nii\ correction)\tablenotemark{a}}
\tablehead{ 
\colhead{\ha\ Line Flux Limit\tablenotemark{b}} &\colhead{$0.8<z<1.2$} & \colhead{$1.85<z<2.2$} & \colhead{$0.7<z<1.5$} & \colhead{$1.5<z<2.0$} & \colhead{$0.7<z<2.0$} }
\startdata
1.0 & $5578^{+485}_{-273}$ & $3730^{+111}_{-1465}$ & $9665^{+873}_{-140}$ & $9813^{+983}_{-2526}$ & $19478^{+1315}_{-2530}$ \\
2.0 & $2015^{+249}_{-263}$ & $976^{+91}_{-539}$ & $3505^{+447}_{-168}$ & $2848^{+60}_{-942}$ & $6353^{+451}_{-939}$ \\
3.0 & $1004^{+159}_{-188}$ & $420^{+68}_{-273}$ & $1769^{+233}_{-97}$ & $1282^{+66}_{-492}$ & $3052^{+242}_{-502}$ \\
5.0 & $369^{+98}_{-97}$ & $136^{+25}_{-99}$ & $672^{+77}_{-87}$ & $440^{+23}_{-216}$ & $1113^{+80}_{-233}$ \\
7.5 & $151^{+61}_{-58}$ & $52^{+14}_{-42}$ & $285^{+49}_{-62}$ & $179^{+14}_{-106}$ & $464^{+50}_{-122}$ \\
10.0 & $76^{+34}_{-35}$ & $25^{+10}_{-21}$ & $147^{+36}_{-43}$ & $91^{+10}_{-58}$ & $238^{+37}_{-72}$
\enddata
\tablenotetext{a}{All counts are per deg$^2$. Fluxes have been corrected for survey incompleteness. The errors account for uncertainties in the best-fit parameters along with the normal Poisson errors.}
\tablenotetext{b}{in $10^{-16}$ ergs s$^{-1}$ cm$^{-2}$.}
\label{tab:counts}
\end{deluxetable*}

\begin{deluxetable*}{cccccc}
\centering
\tablecolumns{6} 
\tablewidth{\textwidth} 
\tablecaption{Cumulative \ha\ number counts (without applying \nii\ correction)\tablenotemark{a}}
\tablehead{ 
\colhead{\ha+\nii\ Line Flux Limit\tablenotemark{b}} &\colhead{$0.8<z<1.2$} & \colhead{$1.85<z<2.2$} & \colhead{$0.7<z<1.5$} & \colhead{$1.5<z<2.0$} & \colhead{$0.7<z<2.0$} }
\startdata
1.0 & $8614^{+823}_{-175}$ & $6736^{+1261}_{-2137}$ & $15001^{+982}_{-643}$ & $16924^{+5461}_{-3175}$ & $31925^{+5548}_{-3239}$ \\
2.0 & $3417^{+292}_{-233}$ & $1922^{+450}_{-737}$ & $5917^{+580}_{-277}$ & $5365^{+953}_{-1298}$ & $11282^{+1116}_{-1328}$ \\
3.0 & $1817^{+112}_{-158}$ & $857^{+211}_{-392}$ & $3166^{+369}_{-208}$ & $2524^{+252}_{-815}$ & $5690^{+446}_{-841}$ \\
5.0 & $733^{+95}_{-258}$ & $291^{+140}_{-169}$ & $1303^{+151}_{-173}$ & $908^{+190}_{-422}$ & $2212^{+243}_{-456}$ \\
7.5 & $323^{+67}_{-93}$ & $117^{+78}_{-77}$ & $591^{+72}_{-100}$ & $384^{+132}_{-211}$ & $975^{+151}_{-234}$ \\
10.0 & $171^{+37}_{-58}$ & $59^{+52}_{-43}$ & $321^{+42}_{-70}$ & $202^{+83}_{-121}$ & $524^{+93}_{-140}$
\enddata
\tablenotetext{a}{All counts are per deg$^2$. Fluxes have been corrected for survey incompleteness. The errors account for uncertainties in the best-fit parameters along with the normal Poisson errors.}
\tablenotetext{b}{in $10^{-16}$ ergs s$^{-1}$ cm$^{-2}$.}
\label{tab:counts_n2}
\end{deluxetable*}

\section{Summary and Conclusions}
\label{sec:conclusions}
Upcoming space based missions will be performing galaxy redshift surveys with the aim of understanding the physical origin of dark energy. The constraints that a given mission will be able to place on the dark energy equation of state parameters depend on the surface density of the used tracers. Both \textit{Euclid} and \textit{WFIRST-AFTA} will be using \ha\ and \oiii\ emitters as tracers of the galaxy population and will focus on the $0.7<z<2$ redshift range.  The precise redshift intervals, however, are still being tuned to maximize the scientific output of these missions. Here, we use the WISP survey to extend on our previous work (focused on \ha\ number counts up to z=1.5) and statistically estimate the number counts of \ha\ emission line galaxies in the full $0.7<z<2$ redshift range.

To this aim, we have measured the bivariate \ha --\oiii\ LLF at $z\sim1$, and showed how, at this redshift, \ha\ number counts can be accurately predicted from the \oiii-only line LF, if the relationship between the \ha\ and \oiii\  luminosities is known.  We find that these two luminosities are broadly correlated, admittedly with a large scatter, that is dominated by different oxygen excitation states and amount of galaxy dust extinction. The large scatter is observed both  in the nearby sample ($z\sim0.25$, from SDSS observations) as well as at $z\sim1$. Moreover, we find no significant evolution in the best-fit \oiii--\ha\ relation in the $\sim$4.5 billion years elapsed between these two epochs. We make the working assumption that the relation continues not to evolve significantly out to redshift $z\sim2$, or, in other words, that any evolution is masked by the large scatter observed in the relation. 

To fit the bivariate LLF model to the data, we  introduced a modified Maximum Likelihood Estimator that allows us to properly account for the uncertainties in the line flux measurement. This modification can change the estimate of the best-fit parameters, particularly for models that vary steeply over small range of luminosities. Our simulations show that the modified MLE improves the accuracy of the recovered best-fit parameters -- especially, when dealing with larger samples, where the measurement uncertainties are more significant than the uncertainty introduces by small number statistics.

We combined the direct measurement of the $z\sim 1-1.5$ \ha\  LF with the $z\sim2$ \ha\ LF determined from the \oiii\ LF and the bivariate LLF information to  provide an estimate of the number of \ha\ emitters expected to be observed down to different line flux limits. Our number count estimates in the full $0.7<z<2$ are approximately 40\% lower than those of \cite{geach10} at the bright flux limits (i.e., for line fluxes above $3.5\times 10^{-16}$ ergs s$^{-1}$ cm$^{-2}$ ), confirming, with twice as many fields and with the full redshift range,  the result of Colbert et al (2013) based on the number of \ha\ emitters up to $z\sim1.5$.  However,  we note that the variation in the number counts obtained from different published works in the literature is substantial at these bright flux levels. This is due to a combination of effects, including the different sample selection techniques, fitting algorithms used to obtain the Schechter parameters, as well the different area/depth combinations of various surveys.

The work and results presented in this paper give us a better understanding of the expected performance from future planned galaxy redshift surveys aiming at constraining the properties of dark energy. This is a significant step toward reducing the uncertainty of figure-of-merit for dark energy for both \textit{Euclid} and \textit{WFIRST-AFTA}. In order to further optimize these planned surveys, more homogenous data are needed. \\

\noindent\textbf{\small ACKNOWLEDGMENTS} 

We thank the referee for providing comments that improved the presentation of the results. Support for HST Programs GO-11696, 12283, 12568, 12902 was provided by NASA through grants from the Space Telescope Science Institute, which is operated by the Association of Universities for Research in Astronomy, Inc., under NASA contract NAS5-26555.

\appendix
\section{A: 1--D Simulations to test the modified MLE}
\label{apnd:sim_1d}

Before applying the modified MLE to deriving the \ha --\oiii\ bivariate LLF for our $0.8<z<1.2$ WISP sample, we test the validity of our modifications to the MLE -- results of which are expected to scale to the bivariate case.

We generate 1000 samples of galaxies distributed according to a known luminosity function. We run two sets of simulations for two different sample sizes: (\textit{i}) small (200 sources per sample), roughly the number of sources in our sample, and (\textit{ii}) large (2000 sources per sample), roughly the number of sources expected to be covered by the end of the WISP survey within the redshift range of interest. The simulated galaxies are assigned the typical uncertainties observed for WISP galaxies at similar luminosities are further randomized within that error-bar. We then fit the simulate samples with both the original and modified MLE techniques.

\begin{figure}[h]
\centering
\includegraphics[width=0.7\textwidth]{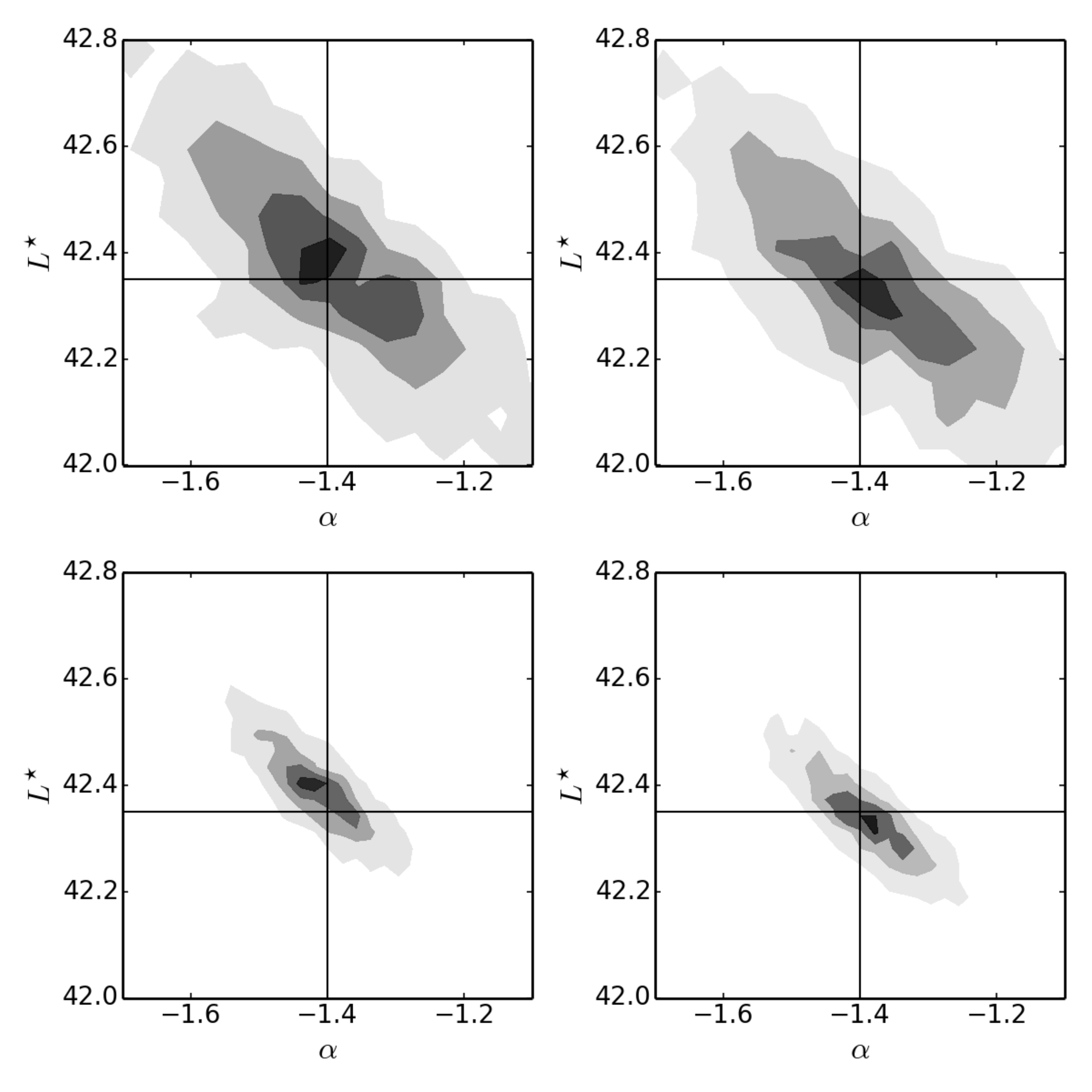}
\caption{Results from the 1-D simulations for small (200 sources per sample; \textit{top row}) and large (2000 sources per sample; \textit{bottom row}) sample sizes comparing the original (\textit{left column}) and modified (\textit{right column}) MLEs. The solid black lines show the true parameters expected to be recovered by the fitting procedures. The contours are at 10\%, 33\%, 66\%, and 95\% confidence levels.} 
\label{fig:sim_1d}
\end{figure}

Figure~\ref{fig:sim_1d} shows the results for the single line LF simulations, for the two different sample sizes for both the original and modified MLE. The modified MLE recovers the true parameters with greater overall accuracy in both large and small sample size cases, even though the scatter is similar. Since measurement uncertainties are not properly treated by the original MLE, a few bright sources with large uncertainties can skew the results significantly. The modified MLE is much less prone to this effect since it marginalizes over the uncertainty. The efficiency of the two estimators depends on what factor is dominating: the statistical randomness of the sample or the measurement uncertainties of the sample.

Since WISP is a slitless grism spectroscopy survey, even the bright sources can have significant uncertainties due to crowding, contamination or other issues. For our sample, the modified MLE is expected to provide an improvement over the original MLE.

\end{document}